\RequirePackage{ifpdf}
\ifpdf 
\documentclass[pdftex]{sigma}
\else
\documentclass{sigma}
\fi

\numberwithin{equation}{section}

\begin{document}

\allowdisplaybreaks

\renewcommand{\thefootnote}{$\star$}

\renewcommand{\PaperNumber}{098}

\FirstPageHeading

\ShortArticleName{Multi-Well Potentials in Quantum Mechanics and Stochastic Processes}

\ArticleName{Multi-Well Potentials in Quantum Mechanics\\ and Stochastic Processes\footnote{This
paper is a contribution to the Proceedings of the Workshop ``Supersymmetric Quantum Mechanics and Spectral Design'' (July 18--30, 2010, Benasque, Spain). The full collection
is available at
\href{http://www.emis.de/journals/SIGMA/SUSYQM2010.html}{http://www.emis.de/journals/SIGMA/SUSYQM2010.html}}}

\Author{Victor P. BEREZOVOJ, Glib I. IVASHKEVYCH and Mikhail I. KONCHATNIJ}

\AuthorNameForHeading{V.P.~Berezovoj, G.I.~Ivashkevych and M.I.~Konchatnij}

\Address{A.I. Akhiezer Institute of Theoretical Physics, National Scientific Center \\ ``Kharkov Institute of Physics and Technology'',
1 Akademicheskaya Str., Kharkov, Ukraine}
\Email{\href{mailto:berezovoj@kipt.kharkov.ua}{berezovoj@kipt.kharkov.ua}, \href{mailto:giv@kipt.kharkov.ua}{giv@kipt.kharkov.ua}, \href{mailto:konchatnij@kipt.kharkov.ua}{konchatnij@kipt.kharkov.ua}}

\ArticleDates{Received October 06, 2010, in f\/inal form December 01, 2010;  Published online December 18, 2010}

\Abstract{Using the formalism of extended $N=4$ supersymmetric quantum
mechanics we consider the procedure of the construction of
multi-well potentials. We demonstrate the form-invariance of
Hamiltonians entering the supermultiplet, using the presented
relation for integrals, which contain fundamental solutions. The
possibility of partial $N=4$ supersymmetry breaking is determined.
We also obtain exact forms of multi-well potentials, both
symmetric and asymmetric, using the Hamiltonian of harmonic
oscillator as initial. The modif\/ication of the shape of potentials
due to variation of parameters is also discussed, as well as
application of the obtained results to the study of tunneling
processes.  We consider the case of exact,
as well as partially broken $N=4$ supersymmetry. The distinctive
feature of obtained probability densities and potentials is
a~parametric freedom, which allows to substantially modify their
shape. We obtain the expressions for probability densities under the generalization of the Ornstein--Uhlenbeck process.}


\Classification{81Q60}

\section{Introduction}

Hamiltonians for systems with multi-well potentials are in the focus of classical and quantum dynamics. One of the most fascinating
applications of multi-well potentials in classical dynamics is the
description of particles transitions from one local minimum to
another under the inf\/luence of dif\/ferent types of noise. This
problem is general for many physical systems from Universe to
microphysics. Calculations of the transition rates from one local minimum to another were initiated by Kramers~\cite{susy-article-sigma-ref:1} and after 70 years of development the problem is far from complete~\cite{susy-article-sigma-ref:2}. From the modern perspective another effect of amplif\/ication of transitions
over the barrier under the action of weak time dependent periodic
signal, which is called stochastic resonance~\cite{susy-article-sigma-ref:3, susy-article-sigma-ref:4} attracts the special attention.

The quantum mechanical dynamics in multi-well potentials is interesting for its own reasons. We just recall the tunneling
ef\/fect, which is in focus again in respect to the
wide-ranging research of trapping of atoms of alkali metals in
superf\/luid state
\cite{susy-article-sigma-ref:5,susy-article-sigma-ref:6}.
Moreover, when more than one barrier exists, the resonant
amplif\/ication of tunneling rate is possible, which is known as
resonant tunneling
\cite{susy-article-sigma-ref:7,susy-article-sigma-ref:8}.
Experimental observation of this phenomenon in superconductive
heterostructures forms  a basis for construction of resonant
tunneling diode. We have to point out the importance of having a mechanism of changing the
parameters of the potential (locations of minima, heights of barriers etc.), to determine the conditions of resonant tunneling.

Classical and quantum systems mentioned in the above, are common in methods of their analysis.
The dynamics of stochastic systems is described by the Fokker--Planck equation
(FP)
\cite{susy-article-sigma-ref:9,susy-article-sigma-ref:10}. One of
the  methods for solving the FP equation is the eigenfunction
expansion method. Due to a~formal similarity of the FP and the Schr\"odinger
equations, the eigenfunction expansion method, which is analogous to the bound
state expansion of the Schr\"odinger equation, can be applied to the Fokker--Planck equation. For instance, in the case
of bistable stochastic system, the knowledge of the full set of wave
functions and eigenvalues of the corresponding quantum Hamiltonian
completely determines the time evolution of the solutions to the  FP
equation. Moreover, it  allows one to make  conclusions about the
dynamics of the corresponding metastable system. Analysis
of   processes in multi-well potentials is complicated due to
the fact that the existing models deal usually with piecewise
potentials (such as constructed from rectangular or parabolic  wells
and barriers), which are presumably far from the real potentials. That is why analytic wave functions and spectrum in such potentials are unknown, which implies the only numerical analysis of their
properties.

Solving the FP equation exactly in this approach is closely
related to the existence of exactly solvable quantum-mechanic
problems, the number of which signif\/icantly increases during the last years
\cite{susy-article-sigma-ref:11,susy-article-sigma-ref:12}. Recall that the f\/irst meaningful example of
construction of the signif\/icantly nonlinear models of dif\/fusion in
bistable system was constructed in~\cite{susy-article-sigma-ref:13,susy-article-sigma-ref:14}. The
formalism of the Darboux transformation, used in these papers, is the one of the basic methods in constructing the
iso\-spectral Hamiltonians in supersymmetric quantum mechanics (SUSY~QM). Further development in the construction of the exactly
solvable stochastic models was achieved in
\cite{susy-article-sigma-ref:15}, where the Fokker--Planck models with
prescribed properties were constructed by use of the Darboux--Grum procedure.
Existence of exactly solvable \cite{susy-article-sigma-ref:16} and
partially solvable \cite{susy-article-sigma-ref:17} quantum
mechanical models with multi-well potentials could f\/ill the gap in
theoretical analysis of the above mentioned processes and may be considered as a
more realistic approximation in their research. Approach based on the accounting the instanton contribution in double-well-like potentials \cite{susy-article-sigma-ref:17a, susy-article-sigma-ref:17b} is commonly used in the description of tunneling processes and, in particular, in studying the features of the Bose--Einstein condencates in multi-well traps. The instanton calculus is also well applied to the transition between the wells under the inf\/luence of noise, because the dynamics of these processes is mainly determined by the energy of the f\/irst excited state.

This paper is aimed at determining the
relations between dif\/ferent types of potentials in the
framework of the extended supersymmetric quantum mechanics ($N =
4$ SUSY~QM) \cite{susy-article-sigma-ref:18,
susy-article-sigma-ref:19,susy-article-sigma-ref:20}. We also study the possibility of changing parameters of potentials in
wide range and obtaining the exact expressions for spectrum and
wave functions. The latter is especially important when potential is equipped with two and even more local minima, so the ``resonant'' tunneling may be realized. The obtained expressions are  used for derivation
of new exactly solvable stochastic models. Here we develop the approach of~\cite{susy-article-sigma-ref:21} to the construction of exactly
solvable stochastic models for potentials with
several local minima.  Contrary to the case of previously established
connection between the FP equation and $N=2$ SUSY~QM, the considered
approach allows one to extend the range of exactly solvable
stochastic models, since the super Hamiltonian of
$N=4$ SUSY~QM contains large number of the isospectral Hamiltonians.
A~special characteristic of the approach is the existence
of a~parametric freedom in probability densities and new
potentials entering the Langevin equation. It makes, in particular,  possible to
change the shape of potentials and densities without changing the
time dependence of the probability density. This fact makes
possible to study the characteristics of stochastic dynamics of
particles in multi-well potentials with variable parameters. On
the other hand, the development of new exactly solvable
dif\/fusional models leads to the construction of more adequate
approximations to real processes.

In Section \ref{section2} we brief\/ly discuss the procedure of
construction of isospectral Hamiltonians with additional states
below the ground state of the initial Hamiltonian in $N=4$ SUSY~QM. We
especially emphasize that multi-well potentials can arise due
to general properties of the solutions to the Schr\"odinger type
auxiliary equation. In particular, we discuss the phenomenon of
partial supersymmetry breaking in $N=4$~SUSY~QM. In the next
section we give general expressions, that allows  one to analyze the
obtained potentials and wave functions, independently on the concrete
form of the initial Hamiltonian. It allows, for instance,  to
generalize the concept of form-invariant potentials
\cite{susy-article-sigma-ref:22} and to calculate the
normalization constants of zero-modes wave functions in the case
of exact and partially broken supersymmetry. Section \ref{section4} is devoted
to the construction of stochastic models, by use of connections in
the supermultiplet of isospectral Hamiltonians of $N=4$~SUSY~QM.
We obtain the expressions for corresponding probability functions
and potentials, entering the Langevin equation. Using the harmonic
oscillator Hamiltonian as the initial one, we obtain the expressions of
the potentials of isospectral Hamiltonians and wave functions for
the case of exact and partially broken supersymmetry. The analysis
of conditions for emergence of double and triple-well potentials
and ways for varying of their form in wide range is provided in
Section~\ref{section5}. The proposed scheme of construction of new
stochastic models is demonstrated in Section~\ref{section6} on the example of
generalization of well known Ornstein--Uhlenbeck process. We give
the results of the calculations of probability densities and
discuss the ways of their modif\/ication when parameters of scheme
vary.  In conclusion we summarize the main results and their
possible applications.

\section[Isospectral Hamiltonians of $N = 4$ SUSY QM with additional states]{Isospectral Hamiltonians of $\boldsymbol{N = 4}$~SUSY~QM\\ with additional states}
\label{section2}

Extended $N = 4$~SUSY~QM \cite{susy-article-sigma-ref:18,
susy-article-sigma-ref:19,susy-article-sigma-ref:20} is equivalent
to the second-order polynomial SUSY QM (reducible
case)~\cite{susy-article-sigma-ref:23,
susy-article-sigma-ref:24,susy-article-sigma-ref:25,susy-article-sigma-ref:26}
and assumes the existence of complex operators of sypersym\-metries
$Q_1$ ($\bar Q_1$) and $Q_2$ ($\bar Q_2$), through which the
Hamiltonians $H_{\sigma_1}^{\sigma_2}$ can be expressed.
Hamiltonian of $N = 4$~SUSY~QM has a form ($\hbar = m = 1$):
\begin{gather*}
  H_{\sigma_1}^{\sigma_2}=\frac{1}{2}\big(p^2+V_2^2(x)+\sigma_3^{(1)}V'_2(x)\big)\equiv \frac{1}{2}\big(p^2+V_1^2(x)+\sigma_3^{(2)}V'_1(x)\big), \nonumber\\
  V_i(x)=W'(x)+\frac{1}{2}\sigma_3^{(i)}\frac{{W''(x)}}{{W'(x)}},
    \end{gather*}
where $W(x)$ is a superpotential and $\sigma_3^{(i)}$ are matrices,
which commute with each other and have eigenvalues $\pm 1$,
$\sigma_3^{(1)}= \sigma_3 \otimes 1$, $\sigma_3^{(2)}= 1 \otimes
\sigma_3$. Supercharges $Q_i$ of extended supersymmetric quantum
mechanics form the algebra:
    \begin{gather}
        \left\{{Q_i, \bar Q_k}\right\}= 2\delta_{ik}H, ~\left\{{Q_i, Q_k}\right\}= \left\{{\bar Q_i, \bar Q_k}\right\}= 0, \qquad i, k = 1, 2,\nonumber\\
        Q_i = \sigma_-^{(i)}(p+iV_{i+1}(x)),  \qquad \bar Q_i = \sigma_+^{(i)}(p - iV_{i+1}(x)),
    \label{susy-article-sigma-eqs:12}
    \end{gather}
where $V_3(x)\equiv V_1(x)$, $\sigma_{_\pm}^{(1)}= \sigma_\pm
\otimes 1$, $\sigma_\pm^{(2)}= 1 \otimes \sigma_\pm$. Hamiltonian
and supercharges act on four-dimensional internal space and
Hamiltonian is diagonal on vectors  $\psi_{\sigma_1}^{\sigma_2}(x,
E)$,  where~$\sigma_1$,~$\sigma_2$ are eigenvalues of
$\sigma_3^{(1)}$, $\sigma_3^{(2)}$. Supercharges $Q_i$ ($\bar Q_i$)
act as lowering (raising) operators for indices~$\sigma_1$,~$\sigma_2$. It is convenient to represent the Hamiltonian structure and the connection between wave functions in diagram form (see below).

\begin{figure}[th]
\centerline{\includegraphics[width=0.5\linewidth]{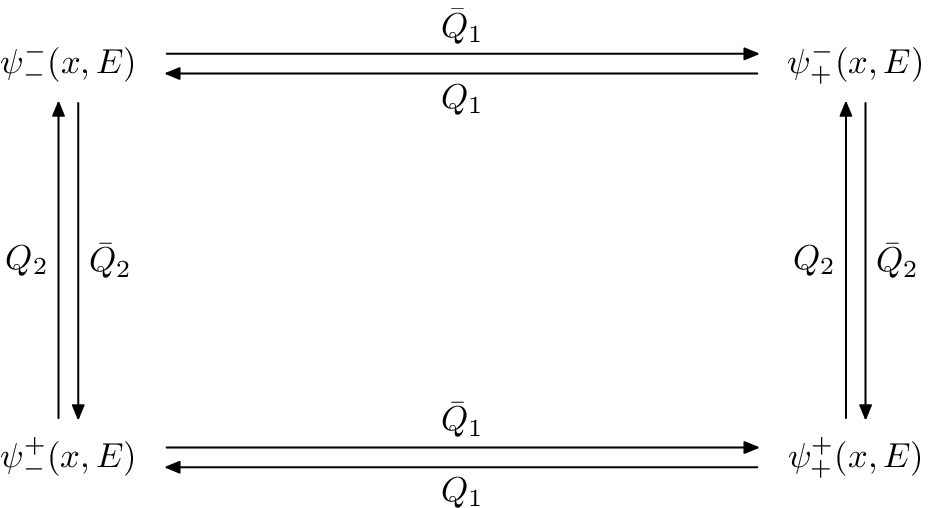}}
\end{figure}

Obviously,  due to commutativity of operators $Q_i$ and $\bar Q_i$
with Hamiltonian, all $\psi_{\sigma_1}^{\sigma_2}(x, E)$ are
eigenfunctions of Hamiltonian with the same eigenvalue $E$. The
only exception is the case, when the wave functions turn to $0$ under
the action of generators of supersymmetry.

Construction of isospectral Hamiltonians within $N =
4$~SUSY~QM is based on the fact that four Hamiltonians are
combined into the supermultiplet $H_{\sigma_1}^{\sigma_2}$.
Nevertheless, it has to be noted that due to the symmetry under $\sigma_1
\leftrightarrow \sigma_2$, i.e.\ $H_{\sigma_1}^{-\sigma_2}\equiv
H_{-\sigma_1}^{\sigma_2}$ only three of them are nontrivial. Let's
note, that such a~relation in the case of higher-derivative  SUSY
\cite{susy-article-sigma-ref:24} establishes the correspondence
between a~quasi-Hamiltonian and operators of the Schr\"odinger type, and
is identical for any superpotentials. The procedure of
construction of isospectral Hamiltonians, when ground state is
removed from the initial Hamiltonian, is considered in~\cite{susy-article-sigma-ref:20} in detail. We will consider the
construction of isospectral Hamiltonians by adding the states
above the ground state of initial Hamiltonian. Similar procedure
was already performed (e.g.\ in \cite{susy-article-sigma-ref:21}),
but the distinctive feature of the present research is to obtain
of general results without specif\/ication of concrete form of
the initial Hamiltonian. Let's consider the auxiliary equation:
    \begin{gather}
        H\varphi(x)= \varepsilon \varphi(x).
    \label{susy-article-sigma-eqs:13}
    \end{gather}
Let's take one of the Hamiltonians as the initial one
    \begin{gather}
         H_{\sigma_1}^{\sigma_2}= \frac{1}{2}(p - i\sigma_1 V_2(x))(p+i\sigma_1 V_2(x))+\varepsilon   \equiv\frac{1}{2}(p - i\sigma_2 V_1(x))(p+i\sigma_2 V_1(x))+\varepsilon,
    \label{susy-article-sigma-eqs:14}
    \end{gather}
where $\varepsilon$ is the so-called factorization energy.
Hereafter the energy is measured from $\varepsilon$. Strictly
speaking, the supersymmetry relations with supercharges (\ref{susy-article-sigma-eqs:12}) expressed through
superpoten\-tial~$W(x)$ are satisf\/ied for shifted by~$\varepsilon$ Hamiltonian $H -
\varepsilon$. However, due to commutativity of supercharges with
constant $\varepsilon$, the relations between wave functions of
$H$ and $H - \varepsilon$ remain the same. When f\/ixing operator
$H_{\sigma_1}^{\sigma_2}$, the form of $W(x)$ depends on the
choice of factorization energy $\varepsilon$, hence the
Hamiltonians $H_{-\sigma_1}^{\sigma_2}$,
$H_{\sigma_1}^{-\sigma_2}$, $H_{-\sigma_1}^{-\sigma_2}$ also have
nontrivial dependence on $\varepsilon$.

When $\varepsilon < E_0$ (where $E_0$ is the ground state energy
of initial Hamiltonian), the auxiliary equation (\ref{susy-article-sigma-eqs:13}) has two linear independent
solutions $\varphi_i(x, \varepsilon)$, $i = 1, 2$, which are
nonnegative and have the following asymptotics
\cite{susy-article-sigma-ref:15}: $\varphi_1(x)\to+\infty$ $(\varphi_2(x)\to 0)$ under $x \to
-\infty$, and
 $\varphi_1(x)\to 0$ $(\varphi_2(x)\to+\infty)$ under $x \to+\infty$,
i.e.\ the general solution has
the form $\varphi(x, \varepsilon, c)= N(\varphi_1(x,
\varepsilon)+c\varphi_2(x, \varepsilon))$ with appropriately chosen constants  ($N$ is the normalization
constant) and has no zeros. Thus, the function
$ \tilde \varphi(x, \varepsilon, c)= \frac{{N^{-
1}}}{{\varphi(x, \varepsilon, c)}}$ is f\/inite and can be
normalized at every concrete choice of~$\varepsilon$ and~$c$.
Let's note that, for some values of~$\varepsilon$ and~$c$,
 $\varphi(x, \varepsilon, c)$ can have local extrema. In this
case the natural choice of the initial Hamiltonian is~$H_-^+$ or~$H_+^-$ (which are identical due to the symmetry of
$H_{\sigma_1}^{\sigma_2}$ under $\sigma_1\leftrightarrow
\sigma_2$). Then the superpotential has the form:
    \begin{gather}
        W(x, \varepsilon, \lambda)= - \frac{1}{2}\ln\left(1+\lambda \int_{x_i}^x{dt}\tilde \varphi^2(t, \varepsilon, c)\right),
    \label{susy-article-sigma-eqs:15}
    \end{gather}
with two new arbitrary parameters $\lambda$, $x_i$, but one of
them is inessential, because it gives an additional contribution
to~$W(x)$. All the Hamiltonians forming the  supermultiplet have
nontrivial dependence on these parameters.

To consider the connection between Hamiltonians from the supermultiplet, let's take $H_+^-$ as the initial one. Denoting  the solution to
    \begin{gather*}
        H_+^-  \psi_+^-(x, E)= E\psi_+^-(x, E)
    \end{gather*}
as $\psi^-_+ (x, E)$ and using the f\/irst representation of the Hamiltonian
$H_{\sigma_1}^{\sigma_2}$  (the l.h.s.\ of~\eqref{susy-article-sigma-eqs:14}), we obtain the following relation
bet\-ween~$H_-^-$, $\psi_-^-(x, E)$ and the initial expressions:
    \begin{gather}
  H_-^- = H_+^-+\frac{{d^2}}{{dx^2}}\ln \tilde \varphi(x, \varepsilon, c),\nonumber\\
        \psi_-^-(x, E_i)= \frac{1}{{\sqrt{2(E_i - \varepsilon)}}}\frac{{W\left\{{\psi_+^-(x, E_i), \varphi(x, \varepsilon, c)}\right\}}}{{\varphi(x, \varepsilon, c)}}, \nonumber\\
      \psi_-^-(x, E = 0)= \frac{{N^{- 1}}}{{\varphi(x, \varepsilon, c)}}= \tilde \varphi(x, \varepsilon, c).
    \label{susy-article-sigma-eqs:17}
    \end{gather}
The new state with $E = 0$ and by def\/inition normalized wave function (remind that energies are measured from~$\varepsilon$) appears in the Hamiltonian.
For the discrete spectrum the normalization of the excited states wave functions is conserved. Using the
second representation (the r.h.s.\ of~\eqref{susy-article-sigma-eqs:14})
\[
 H_{\sigma_1}^{\sigma_2}=
\frac{1}{2}(p - i\sigma_2 V_{\sigma_1}(x))(p+i\sigma_2
V_{\sigma_1}(x))+\varepsilon
\]
 and identity $H_+^- \equiv H_-^+$
the relation between~$H_+^+$, $\psi_+^+(x, E)$ and initial
Hamiltonian and wave functions can be obtained:
    \begin{gather}
  H_+^+= H_+^- + \frac{{d^2}}{{dx^2}}\ln \left({\frac{{\tilde \varphi(x, \varepsilon, c)}}{{1+\lambda \int_{x_i}^x{dt\;\tilde \varphi^2(t, \varepsilon, c})}}}\right) ,\nonumber\\
  \psi_+^+(x, E = 0)= \frac{{N_\lambda^{- 1}\tilde \varphi(x, \varepsilon, c)}}{{ (1+\lambda \int_{x_i}^x{dt\;\tilde \varphi^2(t, \varepsilon, c))}}},
  \nonumber\\
 \psi_+^+(x, E_i)= \frac{1}{{\sqrt{2(E_i - \varepsilon)}}}\left({\frac{d}{{dx}}+\frac{d}{{dx}}\ln \frac{{\tilde \varphi(x, \varepsilon, c)}}{{ (1+\lambda \int_{x_i}^x{dt\;\tilde \varphi^2(t, \varepsilon, c))}}}}\right)\psi_+^-(x, E_i).
     \label{susy-article-sigma-eqs:18}
    \end{gather}
It is worth mentioning that the normalization of the wave function
$\psi_+^+(x, E =0)$, as in the case of one-well potentials, can
always be performed at any $\tilde \varphi(x, \varepsilon, c)$ and
$\lambda$ by use of the following expression in the normalization
condition:
\[
  \frac{{N_\lambda^{- 2}\tilde \varphi^2(x, \varepsilon, c)}}{{(1+\lambda N^{- 2}\int_{- \infty}^x{dt\tilde \varphi^2(t, \varepsilon, c)})^2}}= - \frac{{N_\lambda^{- 2}}}{{\lambda N^{- 2}}}\frac{d}{{dx}}\frac{1}{{(1+\lambda N^{- 2}\int_{- \infty}^x{dt\tilde \varphi^2(t, \varepsilon, c)})}}.
\]
From this relation it is easy to derive the relation between the normalization constants:
$N_\lambda^{-2}=(1+\lambda)N^{-2}$. The normalization of $\psi_+^+(x,
E_i)$ is the same as for $\psi_+^-(x, E_i)$ for any   of
$\tilde \varphi(x, \varepsilon, c)$. The usage of superpotential
(\ref{susy-article-sigma-eqs:15}) with $\tilde \varphi(x,
\varepsilon, c)$ corresponds to exact supersymmetry, which leads
to existence of zero-modes in~$H_-^-$ and~$H_+^+$. Existence of
two zero-modes in super Hamiltonian of $N = 4$~SUSY~QM is caused
by the fact that the Witten index theorem has to be modif\/ied when intertwining
conditions are nonlinear,  as discussed in detail in~\cite{susy-article-sigma-ref:23}.

Let's consider the case when expression
(\ref{susy-article-sigma-eqs:15}) contains one of the particular
solutions, e.g.\ $\varphi_1(x, \varepsilon)$, instead of
$\varphi(x, \varepsilon, c)$. If one of the particular solutions
of second order dif\/ferential equation is known, the second
solution can be obtained from the relation
\[
\varphi_2(x, \varepsilon)= \varphi_1(x,
\varepsilon)\int_{-\infty}^x{dt\frac{1}{{\varphi_1^2(t,
\varepsilon)}}}.
\] Thus, the superpotential
(\ref{susy-article-sigma-eqs:15}) becomes
    \begin{gather*}
  W(x, \varepsilon, \lambda)= - \frac{1}{2}\ln\left(1+\lambda \int_{- \infty}^x{dt}\;\frac{1}{{\varphi_1^2(t, \varepsilon)}}\right)\nonumber \\
\phantom{W(x, \varepsilon, \lambda)}{} \equiv - \frac{1}{2}\ln \left({\frac{{\varphi_1(x, \varepsilon)}}{{\varphi_1(x, \varepsilon)+\lambda \varphi_2(x, \varepsilon)}}}\right)= - \frac{1}{2}\ln \left({\frac{{\varphi_1(x, \varepsilon)}}{{\varphi(x, \varepsilon, \lambda)}}}\right).
    \end{gather*}
It is easy to show, that the state with the energy $E = 0$ in the
spectrum of $H_-^-$ is absent (i.e., the wave function
$\psi_-^-(x, E = 0)$ is nonnormalizable) hence the
spontaneously broken $N = 2$ supersymmetry exists. On the one
hand, the zero energy state has the wave function
\[
\psi_+^+(x, E = 0)\sim \frac{1}{{\varphi(x, \varepsilon,
\lambda)}},
\] normalizable for certain values of
$\lambda > 0$, appears in the spectrum of~$H_+^+$. Here
the exact $N = 2$ supersymmetry takes place. Furthermore,
it is known~\cite{susy-article-sigma-ref:27,
susy-article-sigma-ref:28}, that the partial supersymmetry breaking is
impossible in $N = 4$ SUSY~QM without central
charges. This contradiction resolves with taking into
consideration that the employment of a~factorization energy $\varepsilon
$ in the construction of isospectral Hamiltonians is the simplest way
of incorporation of central charges in $N = 4$ SUSY~QM. The
similar situation occurs in consideration of form-invariant
potentials~\cite{susy-article-sigma-ref:29}. More complete and
consistent consideration of $N = 4$~SUSY~QM with central charges
is given in~\cite{susy-article-sigma-ref:30}, but this point is over
the scope of the current paper.

\section{General properties of multi-well potentials}\label{section3}

We get started the consideration of properties of isospectral
Hamiltanians, derived in previous section without appealing to the form of the initial Hamiltonian, from  the following usefull expressions. Let $y_1$ and $y_2$ be two linear
independent solutions of a homogeneous second order dif\/ferential
equation. Then the following expression   holds
\cite{susy-article-sigma-ref:31}:
    \begin{gather*}
    \int_{x_i}^x\! {W\{y_1, y_2\} \over\left(A_1 y_1(t)\!+\!A_2 y_2(t)\right)^2}dt = - {1\over A_1^2\!+\!A_2^2}\! \left[\left({A_2 y_1(x)\!-\! A_1 y_2(x)\over A_1 y_1(x)\!+\! A_2 y_2(x)}\right)
  - \left({A_2 y_1(x_i)\!-\! A_1 y_2(x_i)\over A_1 y_1(x_i)\!+\!A_2 y_2(x_i)}\right)\right]\!.
    \end{gather*}
Here $W\{y_1, y_2\}= y_1 y'_2 - y'_1y_2$ is the Wronskian, which for the second order dif\/ferential equation,
reduced to canonical form, as the Schr\"odinger equation, is independent on~$x$ and thus
could be taken out of the integral.  This expression is very useful for
calculations of  integrals in expressions (\ref{susy-article-sigma-eqs:15})--(\ref{susy-article-sigma-eqs:18}). First
of all, it is natural to set $x_i = - \infty$ in (\ref{susy-article-sigma-eqs:15}), because
\[
 \tilde \varphi(x, \varepsilon, c )= \frac{{N^{-
1}}}{{\varphi_1(x, \varepsilon)+c \varphi_2(x, \varepsilon)}}
\]
tends to $0$ in the limit and the function with asymptotic
$ {1\over\varphi_i(x, \varepsilon)}\to 0$ under $x \to
- \infty $ always exists, when we use a~particular solution
$\varphi_i(x, \varepsilon)$. Therefore
    \begin{gather}
  N^{- 2}\int_{- \infty}^x{\frac{{dt}}{{\left({\varphi_1(t, \varepsilon)+c\;\varphi_2(t, \varepsilon)}\right)^2}}}
  = - \frac{{N^{- 2}}}{{ (1+c^2) W\{\varphi_1, \varphi_2 \}}}\left[{\Delta(x, \varepsilon, c)- \Delta(- \infty, \varepsilon, c)}\right],\nonumber \\
 N^{- 2}= - \frac{1}{{ (1+c^2) W\{\varphi_1, \varphi_2 \}}}\left[{\Delta(+\infty, \varepsilon, c)- \Delta(- \infty, \varepsilon, c)}\right],\nonumber \\
  \Delta(x, \varepsilon, c)= \frac{{c\varphi_1(x, \varepsilon)- \varphi_2(x, \varepsilon)}}{{\varphi_1(x, \varepsilon)+c\varphi_2(x, \varepsilon)}}.
    \label{susy-article-sigma-eqs:22}
    \end{gather}
Let's f\/ix $c = 1$ in~(\ref{susy-article-sigma-eqs:18}). This  choice allows
one to simplify the consideration, but nevertheless, reveals the
fundamental features of Hamiltonians and wave functions of $N =
4$~SUSY~QM, using only general properties of the solutions to the
auxiliary equation under $\varepsilon < E_0$. \mbox{Using} relations~\eqref{susy-article-sigma-eqs:22}, it is easy to obtain the following modulo constant term expression for the
superpotential:
    \begin{gather}
  W(x, \varepsilon, \lambda)= - \frac{1}{2}\ln \left({1+\lambda N^{- 2}\int_{- \infty}^x{\frac{{dt}}{{\left({\varphi_1(x, \varepsilon)+\varphi_2(x, \varepsilon)}\right)^2}}}}\right) \nonumber\\
\phantom{W(x, \varepsilon, \lambda)}{}
= - \frac{1}{2}\ln \left({\frac{{\varphi_1(x, \varepsilon)+\Lambda(\varepsilon, \lambda)\varphi_2(x, \varepsilon)}}{{\varphi_1(x, \varepsilon)+\varphi_2(x, \varepsilon)}}}\right),\nonumber\\
 \Lambda(\varepsilon, \lambda)= \frac{{\Delta(\infty, \varepsilon,1)- \lambda -(\lambda+1) \Delta(- \infty, \varepsilon,1)}}{{\Delta(\infty, \varepsilon,1)+\lambda -(\lambda+1) \Delta(- \infty, \varepsilon,  1)}}.
        \label{susy-article-sigma-eqs:23}
    \end{gather}
Let's note that $\Delta(\pm\infty, \varepsilon, \lambda)$, entering $\Lambda(\varepsilon, \lambda)$, are determined by the
asymptotics of solutions to the auxiliary equation. Due to
this fact, and since for $H_-^-$ the potential is determined by
the symmetric combination $\varphi_1(x, \varepsilon)+\varphi_2(x,
\varepsilon)$,  while in the case of $H_+^+$ -- by the asymmetric combination
$\varphi_1(x, \varepsilon)+\Lambda(\varepsilon,
\lambda) \varphi_2(x, \varepsilon)$, we get:
    \begin{gather}
        H_-^-= H_+^- - \frac{{d^2}}{{dx^2}}\ln(\varphi_1(x, \varepsilon)+\varphi_2(x, \varepsilon)),
  \nonumber  
\\
        H_+^+= H_+^- - \frac{{d^2}}{{dx^2}}\ln \left({\varphi_1(x, \varepsilon)+\Lambda(\varepsilon, \lambda) \varphi_2(x, \varepsilon)}\right).
    \label{susy-article-sigma-eqs:25}
    \end{gather}

In some sense, the potentials $\bar U_-^-(x, \varepsilon)$ and $\bar U_+^+(x,
\varepsilon, \lambda)$ are form-invariant
\cite{susy-article-sigma-ref:15}, i.e.\ potentials and wave
functions transform to each other by changing of parameters and
their spectra are identical and this holds independently on the
choice of the initial Hamiltonian. As it will be shown in the next
section, if the potential in $H_-^-$ is a~multi-well symmetrical
potential, then in $H_+^+$ it should be asymmetrical. Moreover, varying~$\varepsilon$, as well as~$\lambda$, the form of
 $\bar U_+^+(x, \varepsilon, \lambda)$ can change. Relations
(\ref{susy-article-sigma-eqs:22}) and
(\ref{susy-article-sigma-eqs:23}) are also useful to derive
the exact form of the wave functions $\psi_-^-(x, E)$ and $\psi_+^+(x,
E)$. Thus, the expression for the wave functions $\psi_+^+(x, E)$ comes
from the similar expression for~$\psi_-^-(x, E)$ by the substitution
$\varphi_1(x, \varepsilon)+\varphi_2(x, \varepsilon)\to
\left({\varphi_1(x, \varepsilon)+\Lambda(\varepsilon,
\lambda) \varphi_2(x, \varepsilon)}\right)$. In particular, the
normalization constant for $\psi_+^+(x, E = 0)$ can be
obtained from the corresponding expression for \mbox{$\psi_-^-(x, E = 0)$}.

To demonstrate the possibility of partial supersymmetry breaking
in $N = 4$~SUSY~QM we have to show that the wave function
$\psi_+^+(x, E = 0)\sim \frac{1}{{\varphi(x,
\varepsilon, \lambda)}}$ is normalizable. The value of the normalization
constant can be derived from~(\ref{susy-article-sigma-eqs:22})
under $c = \lambda
> 0$. We will return in what follows to the calculation of this constant for
the concrete form of the initial Hamiltonian. In its turn, the potential in $H_+^+$ has the same form as the corresponding potentials in the case of
exact supersymmetry under replacement $\Lambda(\varepsilon,
\lambda)\to \lambda$.

As it follows from (\ref{susy-article-sigma-eqs:18}) and (\ref{susy-article-sigma-eqs:25}),
the parametric dependence on $\lambda$ is only present in $H_+^+$ and~$\psi_+^+(x,E)$. It should be noted that the value of $\lambda$ is not to be
f\/ixed by the normalization condition for the wave function. Existence of
a~parametric freedom is common in building the isospectral
Hamiltonians, which is based on dif\/ferent versions of the inverse
scattering problem~\cite{susy-article-sigma-ref:32,
susy-article-sigma-ref:33,susy-article-sigma-ref:34}. It is caused by the ambiguity in reconstructing quantum-mechanical
potentials from the spectral data. We will show latter that similar
situation exists also in stochastic mo\-dels. This is unexpectedly enough, since the  potentials of stochastic models have direct physical meaning
in contrast to quantum mechanical potentials.

\section{Probability densities functions}\label{section4}

The FP equation is equivalent to the Langevin equation, however its application in physics is more wide, since it is formulated in more
appropriate language of the probability densities $P^ \pm (x,t;$ $x_0 ,t_0
)$. According to \cite{susy-article-sigma-ref:9,
susy-article-sigma-ref:12} the FP equation has the form:
     \begin{gather*}
       \frac{\partial }{{\partial t}}P^ \pm (x,t;x_0 ,t_0 ) = \frac{D}{2}\frac{{\partial ^2 }}{{\partial x^2 }}P^ \pm (x,t;x_0 ,t_0 ) \mp \frac{\partial }{{\partial x}}V(x)P^ \pm (x,t;x_0 ,t_0 ), \\
       P^ \pm (x,t;x_0 ,t_0 ) = \left\langle {\delta (x - x_0 )} \right\rangle ,\qquad  U_ \pm (x) = \pm \int_0^x {dz}  V(z),
     \end{gather*}
where $U_ \pm (x)$ is the potential entering the Langevin equation. Let's set $t_0 = 0$. The Fokker--Planck equation describes the stochastic dynamics of particles in potentials $U_+(x)$ and $U_-(x) = -U_+(x)$. Substituting
\[
     P^ \pm (x,t;x_0 ,0) = \exp \left\{ { - \frac{1}{D}\left[ {U_ \pm (x) - U_ \pm (x_0 )} \right]} \right\}K_ \pm (x,t)
\]
the FP equation transforms into  the imaginary time Schr\"odinger equation:
    \begin{gather}
 - D\frac{\partial }{{\partial t}}K_ \pm (x,t) = \left\{ { - \frac{{D^2 }}{2}\frac{{\partial ^2 }}{{\partial x^2 }} + \frac{1}{2}\left[ {V^2 (x) \pm D\,V'(x)} \right]} \right\}K_ \pm (x,t),\nonumber \\
  K_ \pm (x,t) = \left\langle {x\left| {\exp \left\{ { - \frac{{tH_ \pm }}{D}} \right\}} \right|x_0 } \right\rangle,\qquad
       H_ \pm = - \frac{{D^2 }}{2}\frac{{\partial ^2 }}{{\partial x^2 }} + \frac{1}{2}\left[ {V^2 (x) \pm D\,V'(x)} \right],
    \label{susy-article-sigma-eqs:31}
    \end{gather}
 in which the dif\/fusion constant $D$ can be treated as the ``Planck constant'', while $H_{\pm}$  has the form of $N=2$~SUSY~QM Hamiltonian. Equations (\ref{susy-article-sigma-eqs:31}) are basic to apply of the eigenfunctions expansion method for construct exactly solvable stochastic models by use of the results of corresponding quantum mechanical problems. Details of this method in the framework of $N=2$ SUSY~QM can be found in \cite{susy-article-sigma-ref:9,susy-article-sigma-ref:12}. Utilization of the extended $N=4$ SUSY~QM formalism gives additional possibilities to construct new exactly-solvable models of stochastic processes, since the Hamiltonian $H_{\sigma _1}^{\sigma _2}$ of $N=4$ SUSY~QM includes four isospectral Hamiltonians. It allows to obtain new $K_{\sigma _1 }^{\sigma _2 }(x,t)$ and $U_{\sigma _1 }^{\sigma _2 }(x)$, hence $P_{\sigma _1 }^{\sigma _2 } (x,t;x_0 )$.

To obtain the expressions for probability densities $P_{\sigma _1
}^{\sigma _2 } (x,t;x_0 )$ we should to know not only the wave functions
and the spectrum of $H_{\sigma _1 }^{\sigma _2 } $, but those of the
corresponding $U_{\sigma _1 }^{\sigma _2 } (x)$. It is easy to see
that
    \begin{gather}
      U_ + ^ - (x,c = 1) = - D\ln \left( {\varphi _1 (x,\varepsilon ) + \varphi _2 (x,\varepsilon )} \right), \qquad U_ - ^ - (x,c = 1) = - U_ + ^ - (x,c = 1).
      \label{susy-article-sigma-eqs:32}
    \end{gather}

Further consideration is based on account of the symmetry of
$N=4$~SUSY~QM
$H_{\sigma _1 }^{\sigma _2 }$ under $\sigma _1 \leftrightarrow
\sigma _2$, that leads to the relation:
    \begin{gather}
      H_ + ^ - (x,p) = \frac{1}{2}\bar Q_1^{( - )} Q_1^{( - )} \equiv \frac{1}{2}Q_2^{( + )} \bar Q_2^{( + )} = H_ - ^ + (x,p).
      \label{susy-article-sigma-eqs:33}
    \end{gather}

First equality indicates the existence of the expression of
$H^{(+)}$ in terms of $Q_1$ $(\bar Q_1)$, while the second equality
implies that $H^{(-)}$ is expressed in terms of $Q_2$ $(\bar Q_2 )$. At the
same time, the   supercharges entering $H^{(+)}$ and $H^{(-)}$ are substantially
dif\/ferent:
    \begin{gather}
  H_ - ^ + (x,p) = \frac{1}{2}Q_2^{( + )} \bar Q_2^{( + )} = \frac{1}{2}\big[ {p^2 + \big( {V_1^{( + )} (x)} \big)^2 - DV_1^{( + )}{}^\prime (x)} \big],
  \nonumber \\
  V_1^{( + )} (x) = D\frac{d}{{dx}}\ln \left| {\frac{{\tilde \varphi (x,\varepsilon ,c)}}{{1 + \lambda \int_{x_i }^x {dx'\left[ {\tilde \varphi (x',\varepsilon ,c)} \right]^2 } }}} \right|
  = - D\frac{d}{{dx}}\ln \left| {\varphi _1 (x,\varepsilon ) + \Lambda (\varepsilon ,\lambda )\varphi _2 (x,\varepsilon )} \right| .\!\!
      \label{susy-article-sigma-eqs:34}
    \end{gather}

According to (\ref{susy-article-sigma-eqs:33}),
 quantum $H_-^+$ and $H_+^-$ have the same spectrum and the wave functions,
thus the corresponding $K(x,t)$ are identical. But the
corresponding stochastic models are described by essentially
dif\/ferent potentials, therefore their $P(x,t;x_0 )$ are dif\/ferent.
Moreover, the potential $U_ - ^ + (x,\varepsilon ,\Lambda ) = - U_ + ^
+ (x,\varepsilon ,\Lambda ) = - D\ln \left| {\varphi _1
(x,\varepsilon ) + \Lambda (\varepsilon ,\lambda )\;\varphi _2
(x,\varepsilon )} \right|$ has non-trivial parametric dependence
on $\lambda$, which implies the existence of the family of
stochastic models with the same time dependence of the probability
density, but substantially dif\/ferent coordinates dependence.
Surprisingly, the parametric freedom allows for a modif\/ication of the shape
of potential. Recall that physical quantities, e.g.\ time of
passing the potential maximum, considerably depend on the local
modif\/ication of the  potential form~\cite{susy-article-sigma-ref:35,
susy-article-sigma-ref:36,susy-article-sigma-ref:37}.

It is easy to obtain the corresponding $P_{\sigma _1 }^{\sigma _2 }
(x,t;x_0 )$ using the expressions for wave functions $\psi
_{\sigma _1 }^{\sigma _2 } (x,E_n )$ and potentials $U_{\sigma _1
}^{\sigma _2 } (x)$ from
(\ref{susy-article-sigma-eqs:17}), (\ref{susy-article-sigma-eqs:18})
and
(\ref{susy-article-sigma-eqs:32}), (\ref{susy-article-sigma-eqs:34}).
It is important to note, that transitional probability densities
$P_ - ^ + (x,t;x_0 )$ and $P_ + ^ + (x,t;x_0 )$ have also the
parametric dependence on $\lambda $.

Denoting $\varphi (x,\varepsilon ,c=1) = \varphi _1 (x,\varepsilon
) + \varphi _2 (x,\varepsilon )$, the expression for calculation of
distribution function $P_ - ^ - (x,t;x_0 )$ takes the form:
    \begin{gather}
         P_ - ^ - (x,t;x_0 ) = \frac{{N^{ - 2} }}{{\left( {\varphi (x,\varepsilon ,1)} \right)^2 }} + \left( {\frac{{\varphi (x_0 ,\varepsilon ,1)}}{{\varphi (x,\varepsilon ,1)}}} \right)
  \sum\limits_{n = 0}^N {e^{ - \frac{t}{D}(E_n - \varepsilon )} } \psi _ - ^ - (x,E_n )\;\psi _ - ^ - (x_0 ,E_n ).
      \label{susy-article-sigma-eqs:35}
    \end{gather}
For simplicity let's assume that the spectrum of the initial Hamiltonian
has only the discrete states. When the continuous spectrum exists the
summation should be replaced with integration over the corresponding
density of states. Substituting the expressions for wave functions
$\psi _ - ^ - (x,E_n )$ (\ref{susy-article-sigma-eqs:17}) into \eqref{susy-article-sigma-eqs:35} we
obtain:
    \begin{gather}
        P_ - ^ - (x,t;x_0 ) = \frac{{N^{ - 2} }}{{\left( {\varphi (x,\varepsilon ,1)} \right)^2 }} + \frac{1}{2}\left( {\varphi (x_0 ,\varepsilon ,1)} \right)^2\nonumber \\
\phantom{P_ - ^ - (x,t;x_0 ) =}{}
\times \sum\limits_{n = 0}^N {\frac{{e^{ - \frac{t}{D}(E_n - \varepsilon )} }}{{(E_n - \varepsilon )}}} \frac{{d^2 }}{{dxdx_0 }}\left( {\frac{{\psi _ + ^ - (x,E_n )}}{{\varphi (x,\varepsilon ,1)}}} \right)\left( {\frac{{\psi _ + ^ - (x_0 ,E_n )}}{{\varphi (x_0 ,\varepsilon ,1)}}} \right).
      \label{susy-article-sigma-eqs:36}
    \end{gather}
The expression for $P_ + ^ + (x,t;x_0 )$ can be obtained from
equation (\ref{susy-article-sigma-eqs:36}) by substituting $\varphi
(x,\varepsilon ,1) \to \varphi (x,\varepsilon ,\Lambda (\lambda
,\varepsilon ))$, $N^{ - 2} \to N_\Lambda ^{ - 2} $. Both of $P_ - ^
- (x,t;x_0 )$ and $P_ + ^ + (x,t;x_0 )$ have the equilibrium
distributions at $t \to \infty $, which coincide with square of
 the zero modes  wave functions in Hamilto\-nians~$H_ - ^ - $ and~$H_ +
^ + $. Moreover, it is easy to see that terms from
the excited states of $H_ - ^ - $ and $H_ + ^ + $ do not contribute to
the normalization condition of these density functions, i.e.\ the
normalization of these functions is guaranteed by their equilibrium
values. Dependence on parameter~$\lambda $ in~$P_ + ^ +
(x,t;x_0 )$ is not eliminated by the normalization condition.

The distribution $P_ + ^ - (x,t;x_0 )$ ($P_ - ^ + (x,t;x_0 )$)
describes the stochastic dynamics of a~particle in the metastable state of
the potential, inverted to $U_ + ^ - (x,\varepsilon )$ ($ - U_ + ^
+ (x,\varepsilon ,\Lambda (\varepsilon ,\lambda ))$. Thus, it has
not the equilibrium limit and is determined by the expression:
      \begin{gather}
        P_ + ^ - (x,t;x_0 ) = \frac{{\varphi (x,\varepsilon ,1)}}{{\varphi (x_0 ,\varepsilon ,1)}}
        \sum\limits_{n = 0}^N {e^{ - \frac{{(E_n - \varepsilon )t}}{D}} } \psi _ + ^ - (x,E_n )\psi _ + ^ - (x_0 ,E_n ).
        \label{susy-article-sigma-eqs:37}
      \end{gather}
Similarly to the previous case $P_ - ^ + (x,t;x_0 )$ can be obtained
from $P_ + ^ - (x,t;x_0 )$ by substitution $\varphi (x,\varepsilon
,1)$ $\to \varphi (x,\varepsilon ,\Lambda (\lambda ,\varepsilon ))$.
The parametric dependence on $\lambda $ can be eliminated by
the normalization condition for $P_ - ^ + (x,t;x_0 )$, but as it will
be demonstrated below,  choosing  the initial Hamiltonian of harmonic
oscillator, the parametric arbitrariness in $P_ - ^ +
(x,t;x_0 )$ remains.

To summarize, we should note that in the framework of $N=4$~SUSY~QM
four models of stochastic dynamics emerge. Two of them correspond
to the motion of particle in double-well potentials (symmetric and
asymmetric) under the action of the Gaussian noise, the other ones
describe the dynamics of the particle in the metastable state, when  $N=4$
supersymmetry is exact. For the case of the partial supersymmetry breaking $P_
- ^ - (x,t;x_0 )$, $P_ + ^ - (x,t;x_0 )$ and $P_ - ^ + (x,t;x_0 )$
describe stochastic processes in the metastable state, while $P_ + ^ +
(x,t;x_0)$ -- in bistable. Some of them possess the parametric freedom,
that allows to change the parameters of external f\/ield.

\section{Isospectral Hamiltonians with almost equidistant spectrum}\label{section5}

To construct the explicit expressions for potentials and wave functions we   choose the  initial Hamiltonian with the harmonic oscillator ($HO$) potential. Let's consider the solution to the auxiliary equation for $\varepsilon<E_0 = \frac{\omega}{2}$, $(\hbar=m=1)$:
    \begin{gather*}
        \left({\frac{{d^2}}{{dx^2}}+2\left(\varepsilon - \frac{{\omega^2 x^2}}{2}\right)}\right)\varphi(x, \varepsilon)= 0.
    \end{gather*}
Introducing dimensionless variables $\xi = \sqrt{2\omega}x$ we obtain the equation for $\varphi(\xi, \bar\varepsilon)$, where $ \bar \varepsilon = \frac{\varepsilon}{\omega}$:
    \begin{gather*}
        \left({\frac{{d^2}}{{d\xi^2}}+\left(\nu+\frac{1}{2}- \frac{{\xi^2}}{4}\right)}\right) \varphi(\xi, \bar \varepsilon)= 0 , \qquad \nu = - \frac{1}{2}+\bar \varepsilon.
    \end{gather*}

Now we consider two dif\/ferent cases.

\begin{figure}[t]
\centering
\begin{minipage}{2.6in}
 \includegraphics[width=2.5in]{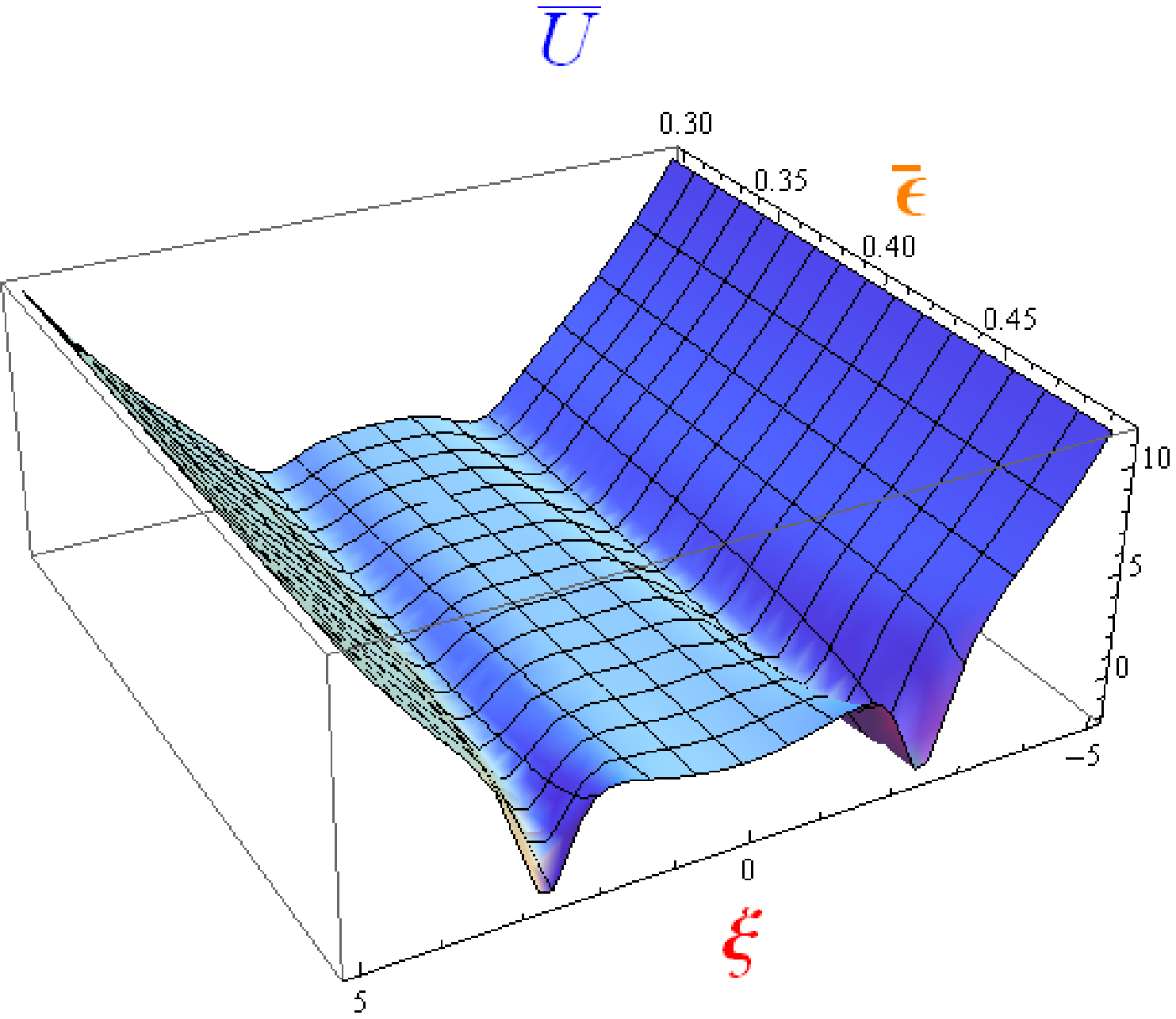}
  \caption{Shape of the potential $\bar U_-^-(\xi, \bar \varepsilon)$.\label{susy-article-sigma-fig:1}}
\end{minipage}\qquad \qquad
\begin{minipage}{2.6in}
\includegraphics[width=2.5in]{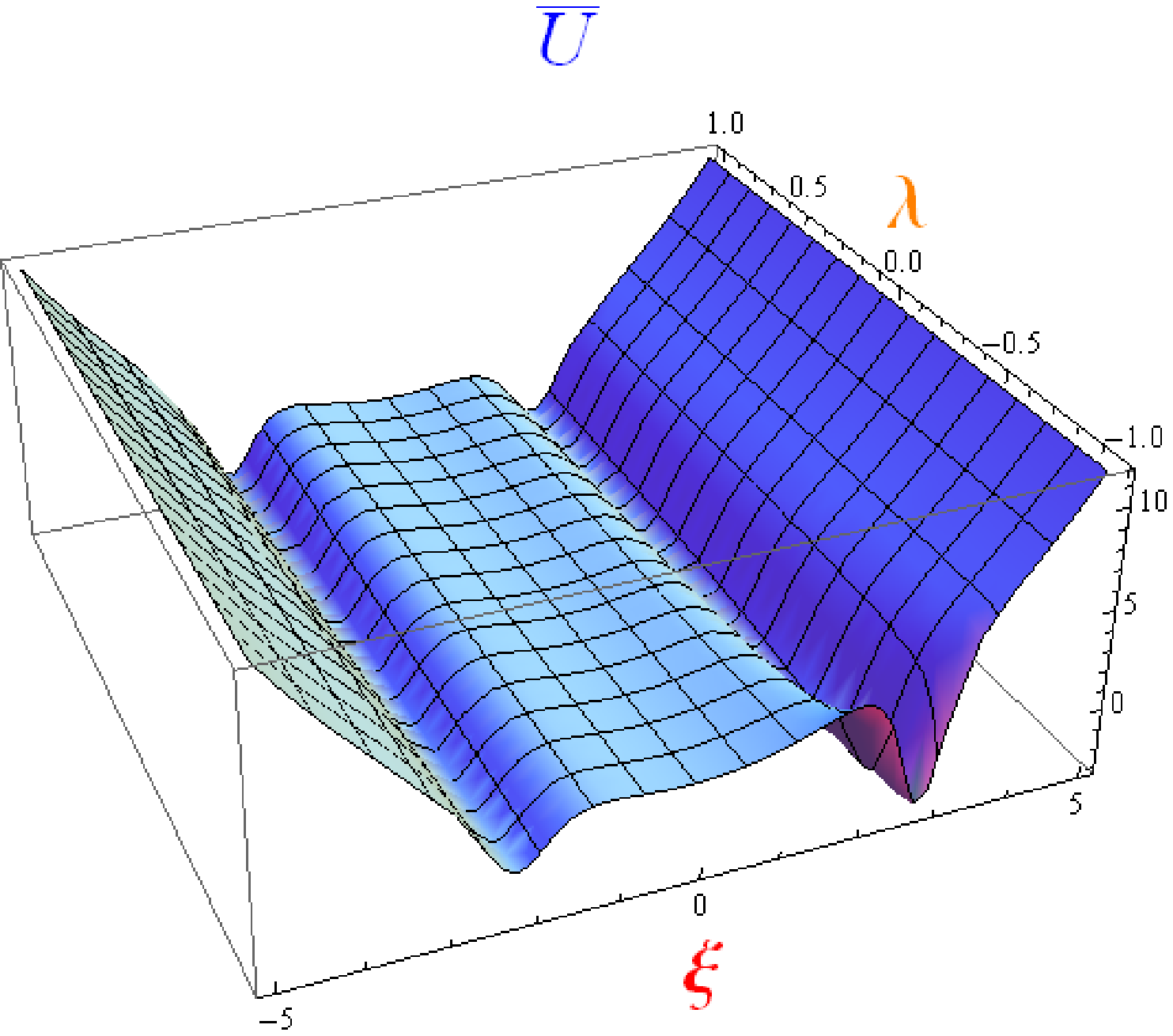}
  \caption{Potential $\bar U_+^+(\xi, \bar \varepsilon = 0.47, \lambda)$.\label{susy-article-sigma-fig:2}}
\end{minipage}
\end{figure}

\textbf{a)} \textbf{Exact $\boldsymbol{N=4}$~SUSY.}
This equation has two linear independent solutions: parabolic cylinder functions $D_\nu(\sqrt 2 \xi)$, $D_\nu(-\sqrt 2 \xi)$. According to terminology in the  above, we denote $\varphi_1(\xi, \bar\varepsilon)= D_\nu(\sqrt 2 \xi)$, $\varphi_2(\xi, \bar\varepsilon)= D_\nu(-\sqrt 2 \xi)$ and the Wronskian  becomes  $ W\{\varphi_1, \varphi_2 \}= \frac{{2\sqrt \pi}}{{\Gamma(- \nu)}}$ \cite{susy-article-sigma-ref:38}, with gamma-function $\Gamma(-\nu)$. Following the procedure from the previous section, the general solution to the auxiliary equation is chosen to be:
    \begin{gather}
        \varphi(\xi, \bar \varepsilon, 1)= D_\nu\big(\sqrt 2 \xi\big)+D_\nu\big(- \sqrt 2 \xi\big).
    \label{susy-article-sigma-eqs:43}
    \end{gather}
As one can see from (\ref{susy-article-sigma-eqs:43}), $\varphi(x, \bar \varepsilon, 1)$ is an even function of $\xi$. To obtain the exact form of the superpotential, the integral, entering the def\/inition of the $W(x, \varepsilon, \lambda)$ as well as the normalization constant~$N^{-2}$ have to be calculated from (\ref{susy-article-sigma-eqs:22}), (\ref{susy-article-sigma-eqs:23}). Due to the symmetry of $\varphi(\xi, \bar\varepsilon)$, expression of the integral simplif\/ies and takes the form:
    \begin{gather*}
          1+\lambda N^{- 2}\int_{- \infty}^\xi   {\frac{{dt}}{{\left({\varphi_1+\varphi_2}\right)^2}}}= 1 - \frac{{\lambda N^{- 2}}}{{2W\{\varphi_1, \varphi_2 \}}} \left[{\left({\frac{{\varphi_1(\xi, \bar \varepsilon) -  \varphi_2(\xi, \bar \varepsilon)}}{{\varphi_1(\xi, \bar \varepsilon) + \varphi_2(\xi, \bar \varepsilon)}}}\right)+\Delta(+\infty, \bar \varepsilon, 1)}\right].
    \end{gather*}
Hence, the superpotential is:
    \begin{gather}
 W(\xi, \bar \varepsilon, \lambda )= \ln \left({\frac{{(1+\frac{\lambda}{2}+\frac{\lambda}{{2\Delta(+\infty, \bar \varepsilon, 1)}})\varphi_1+(1+\frac{\lambda}{2}- \frac{\lambda}{{2\Delta(+\infty, \bar \varepsilon, 1)}})\varphi_2}}{{\varphi_1+\varphi_2}}}\right),
    \label{susy-article-sigma-eqs:45}
\\
\nonumber  N^{- 2}= - \frac{{W\{\varphi_1, \varphi_2 \}}}{{\Delta(+\infty, \bar \varepsilon, 1)}}= \frac{{2\sqrt \pi}}{{\Gamma(- \nu)}}.
\end{gather}
Using the asymptotic value of the parabolic cylinder function, we have $\Delta(+\infty, \bar\varepsilon, 1)=-1$ and the expression for the superpotential simplif\/ies:
    \begin{gather}
          W(\xi, \bar \varepsilon, \lambda)= - \frac{1}{2}\ln \left({\frac{{\varphi_1(\xi, \bar \varepsilon)+(1+\lambda)\varphi_2(\xi, \bar \varepsilon)}}{{\varphi_1(\xi, \bar \varepsilon)+\varphi_2(\xi, \bar \varepsilon)}}}\right)= - \frac{1}{2}\ln \left({\frac{{\varphi(\xi, \bar \varepsilon, 1+\lambda)}}{{\varphi(\xi, \bar \varepsilon, 1)}}}\right).
    \label{susy-article-sigma-eqs:46}
    \end{gather}
From (\ref{susy-article-sigma-eqs:46}) it follows, that the value of $\lambda$ is restricted to $\lambda>-1$. Using~(\ref{susy-article-sigma-eqs:17}) for $H_-^-$ and $\psi_-^-(x, E)$, we can derive the  form of the Hamiltonian  and the corresponding wave functions (here $\psi_+^-(\xi, E_i)$ are the wave functions of the HO):
    \begin{gather}
  H_-^- = H_+^- - \frac{{d^2}}{{dx^2}}\ln \left({D_\nu(\sqrt 2 \xi)+D_\nu(- \sqrt 2 \xi)}\right),\nonumber\\
  \psi_-^-(\xi, E_i)= \frac{1}{{\sqrt{2(E_i - \bar \varepsilon)}}}\frac{{W\left\{{\psi_+^-(\xi, E_i), \varphi(\xi, \bar \varepsilon, 1)}\right\}}}{{\varphi(\xi, \bar \varepsilon, 1)}},\nonumber\\
 \psi_-^-(x, E = 0)= \frac{{N^{- 1}}}{{\left({D_\nu(\sqrt 2 \xi)+D_\nu(- \sqrt 2 \xi)}\right)}}= \frac{{N^{- 1}}}{{\varphi(x, \bar \varepsilon, 1)}}, \qquad  N^{- 2}= \frac{{2\sqrt \pi}}{{\Gamma(- \nu)}}.
    \label{susy-article-sigma-eqs:47}
    \end{gather}
Analysis of (\ref{susy-article-sigma-eqs:47}) reveals that  there exist several local minima   only for $0<\bar\varepsilon<\frac{1}{2}$. With values of $\bar\varepsilon$ to be close to the right boundary, the third local minimum appears (Fig.~\ref{susy-article-sigma-fig:1}) and depth of the outside minima increases. It should be noted that in  terms of the dimensionless variable $\xi$ the only way to vary the form of the potential is by varying $\bar\varepsilon$ and $\lambda$. In the case of natural units, additionally, the form of the potential (in particular, positions of the local minima) can be changed by variation of $\omega$.

Connection between $H_+^+$, $\psi_+^+(\xi, E_n)$ and $H_+^-$,
$\psi_+^-(\xi, E_n)$  can be viewed in the same
manner. Expressions for $H_+^+$ and $\psi_+^+(\xi, E_n)$ can be
obtained from (\ref{susy-article-sigma-eqs:47}) by substituting
$\varphi(\xi,\bar \varepsilon, 1)\rightarrow\varphi(\xi,\bar
\varepsilon, \lambda+1)$:
    \begin{gather*}
         H_+^+= H_+^- - \frac{{d^2}}{{dx^2}}\ln \left({D_\nu(\sqrt 2 \xi)+(1+\lambda)D_\nu(- \sqrt 2 \xi)}\right),\nonumber\\
    \psi_+^+(\xi, E_i)= \frac{1}{{\sqrt{2(E_i - \bar \varepsilon)}}}\frac{{W\left\{{\psi_+^-(\xi, E_i), \varphi(\xi, \bar \varepsilon, \lambda+1)}\right\}}}{{\varphi(\xi, \bar \varepsilon, \lambda+1)}},\nonumber\\
  \psi_-^-(\xi, E = 0)= \frac{{N_{\lambda+1}^{- 1}}}{{\left({D_\nu(\sqrt 2 \xi)+(\lambda+1)D_\nu(- \sqrt 2 \xi)}\right)}}= \frac{{N_{\lambda+1}^{- 1}}}{{\varphi(x, \bar \varepsilon, \lambda+1)}},\nonumber\\
 N_{\lambda+1}^{- 2}= \frac{{2(\lambda+1)\sqrt \pi}}{{\Gamma(- \nu)}}.
        \end{gather*}

From these relations  it follows the restriction $-1< \lambda$. As it can be seen from Fig.~\ref{susy-article-sigma-fig:2}, the potential $\bar U_+^+(\xi, \bar\varepsilon, \lambda)$ possesses the well indicated asymmetry, which increases under $\lambda \to -1$. The value $\bar\varepsilon = 0.47$ corresponds to the region, where $\bar U_+^+(\xi, \bar\varepsilon, \lambda)$ has three local minima. It should be noted, that the depth of the central minima also increases with $\lambda\to -1$. The obtained potentials and the corresponding wave functions may be applied for the resonant tunneling phenomenon~\cite{susy-article-sigma-ref:4} studies.

\begin{figure}[t]
\centering
 \includegraphics[width=0.35\linewidth]{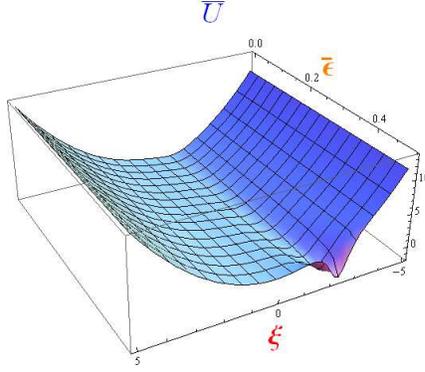}
  \caption{Potential $\bar U_-^-(\xi, \bar \varepsilon)$ ($0 < \bar \varepsilon < 0.5$).
  \label{susy-article-sigma-fig:3}}
\end{figure}

\textbf{b)} \textbf{Partial $\boldsymbol{N=4}$~SUSY  breaking.}
The utilization of the particular solution of auxiliary equation for derivation of superpotential (\ref{susy-article-sigma-eqs:45}) realizes the situation of partial supersymmetry breaking. As it was mentioned above, the spectrum of $H_-^-$ does not contain states with $E = 0$, because their wave function is nonnormalizable. The shape of $\bar U_-^-(\xi, \bar\varepsilon)$ is presented in Fig.~\ref{susy-article-sigma-fig:3}. At the same time, zero state appears in the spectrum of $H_+^+$ with the wave function:
    \begin{gather*}
        \psi_+^+(\xi, E = 0)= \frac{{N_\lambda^{- 1}}}{{\varphi(\xi, \bar \varepsilon, \lambda)}}= \frac{{N_\lambda^{- 1}}}{{\left({\varphi_1(\xi, \bar \varepsilon)+\lambda \varphi_2(\xi, \bar \varepsilon)}\right)}}.
    \end{gather*}
The normalization constant $N_\lambda^{-2}$ is calculated by use of
(\ref{susy-article-sigma-eqs:47}) and has the form $N_\lambda^{-2}= \frac{{2\lambda \sqrt \pi}}{{\Gamma(- \nu)}}$.
Thus, it is restricted to $\lambda > 0$. The Hamiltonian becomes
$ H_+^+=H_+^- -
\frac{{d^2}}{{d\xi^2}}\ln\left({\varphi(\xi, \bar\varepsilon,
\lambda}\right)$. The potential shape ($\bar\varepsilon = 0.47$) is presented
in Fig.~\ref{susy-article-sigma-fig:2} with $\lambda > 0$.

When $\bar \varepsilon = - \frac{1}{2}$ ($D_{- 1}(\sqrt 2 \xi)= e^{\frac{{\xi^2}}{2}}\sqrt{\frac{\pi}{2}}\left({1 - \Phi(\xi)}\right)$, with $\Phi(\xi)$ to be the error function), potentials $\bar U_-^-(x, \bar\varepsilon)$ correspond  to that of~\cite{susy-article-sigma-ref:39} (the only dif\/ference is in additional constant term)
    \begin{gather}
        \bar U_-^-\left(\xi, \bar \varepsilon = - \tfrac{1}{2}\right)= \left({\frac{{\xi^2 - 1}}{2}}\right)+\frac{4}{{\sqrt \pi}}  \frac{{e^{- \xi^2}}}{{\left({1 - \Phi(\xi)}\right)}}\left[{\frac{{e^{- \xi^2}}}{{\sqrt \pi \left({1 - \Phi(\xi)}\right)}}- \xi}\right].
    \label{susy-article-sigma-eqs:410}
    \end{gather}
This potential is single-well, as well as corresponding $ \bar U_+^+(\xi, \bar\varepsilon = -\tfrac 12, \lambda)$, because existence of several local minima is possible only for $0 < \bar \varepsilon < \frac{1}{2}$. Using the form-invariance property of~$\bar U_+^+$:
    \begin{gather}
        \bar U_{+}^+\left(\xi, \bar \varepsilon = -\tfrac 12, \lambda\right)= \frac{{\xi^2 - 1}}{2}- \frac{{d^2}}{{d\xi^2}}\ln \left({(1+\lambda)-(1 - \lambda)\Phi(\xi)}\right).
    \label{susy-article-sigma-eqs:411}
    \end{gather}
The spectrum of the Hamiltonian with (\ref{susy-article-sigma-eqs:411}) contains, in contrast to (\ref{susy-article-sigma-eqs:410}), the state with $E = 0$ and the wave function of the form:
    \begin{gather*}
          \psi_+^+(\xi, E = 0)= \frac{{\left({2\lambda \sqrt \pi}\right)^{1/2}e^{- \frac{{\xi^2}}{2}}}}{{\left({(1+\lambda)-(1 - \lambda)\Phi(\xi)}\right)}}.
    \end{gather*}

\section[Generalization of the Ornstein-Uhlenbeck process]{Generalization of the Ornstein--Uhlenbeck process}\label{section6}

Let us give an example of the construction of new stochastic models. We choose the harmonic oscillator (HO) Hamiltonian   as the initial one. Then, we consider the auxiliary equation with $\varepsilon < E_0 = \frac{{D\omega}}{2}$ (recall, that $D$ plays the role of the ``Planck constant'' in the FP equation and $\bar \varepsilon = {\varepsilon\over D\omega}$). Here we consider two cases.

\textbf{a)} \textbf{Exact $\boldsymbol{N=4}$~SUSY.}
For many problems of stochastic dynamics it is reasonable to use
symmetric double-well potentials, so we consider  f\/irst the case
$c = 1$. As it can be seen from~(\ref{susy-article-sigma-eqs:43}), $\varphi (\xi ,\bar \varepsilon
,1)$ is even function, and the normalization constant $N^{ -
2}$ can be calculated from~(\ref{susy-article-sigma-eqs:45}).
One of the main characteristics of the stochastic process is the
potential entering the Langevin equation. As it follows from~(\ref{susy-article-sigma-eqs:32}), $U_-^-(x) = - U_ + ^ - (x)
= D\ln \varphi (x,\bar \varepsilon, 1)$ is symmetric, while  $U_ +
^ + (x,\Lambda) = - U_ - ^ + (x, \Lambda)= D\ln \varphi (x,\bar
\varepsilon,\lambda+1)$ is asymmetric, that depends on the value of~$\lambda$. The existence of the parametric freedom in $U_+^+(x,\Lambda)$
is quite surprising: contrary to quantum mechanics, where
potentials reconstructed from the spectral data indeed have a~parametric
freedom, but give the same observables, in stochastic mechanics
local properties of the potential substantially inf\/luence on many
characteristics, for example, on the rates of the potential barrier
crossing~\cite{susy-article-sigma-ref:35,susy-article-sigma-ref:36,susy-article-sigma-ref:37}.

\begin{figure}[t]\centering
      \begin{tabular}{ccc}
      a) \includegraphics[width=0.35\linewidth]{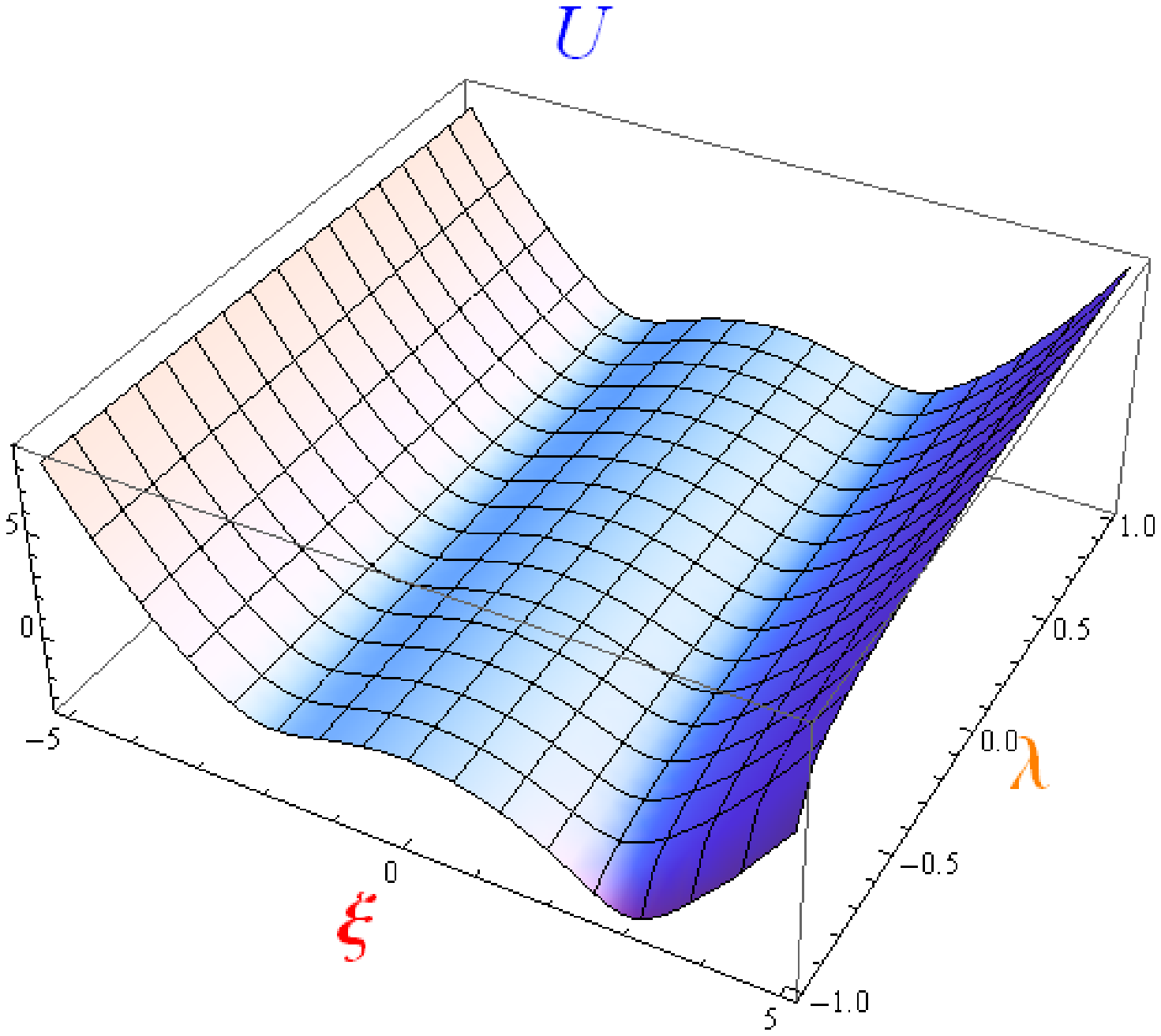}
      &\qquad\qquad &
      b) \includegraphics[width=0.35\linewidth]{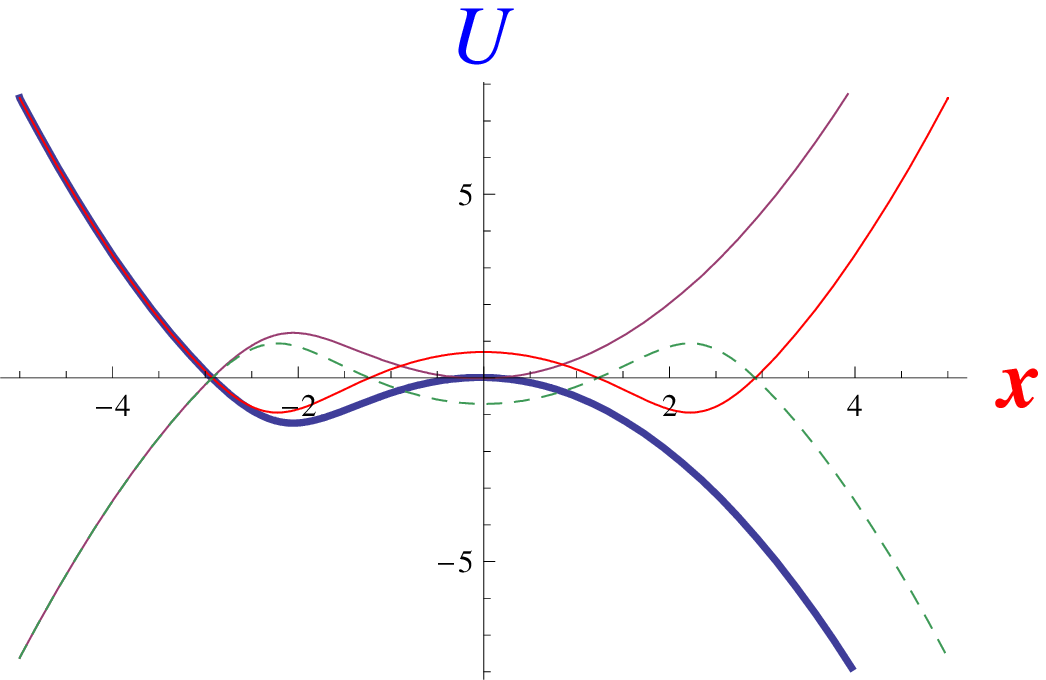}
      \end{tabular}
      \caption{Potentials $U_{\sigma _1 }^{\sigma _2 }$.
      a) exact $N=4$~SUSY: $U_ + ^ + (\xi ,\lambda )$ ($\omega = D = 1$, $\bar \varepsilon = 0.47$); b) Partial broken $N=4$~SUSY: $U_ + ^ - (\xi )$ -- lilac, $U_ - ^ - (\xi )$ -- blue, $U_ - ^ + (\xi,\lambda )$ -- dashed, $U_ + ^ + (\xi ,\lambda )$ -- red ($\bar \varepsilon = 0.47$, $\lambda = 1$).\label{susy-article-all-fig:1}}
      \end{figure}

$U_ - ^ + (x,\Lambda ) = - U_ + ^ + (x,\Lambda ) = -
D\ln \left| {\varphi _1 (x,\bar \varepsilon ) + (\lambda+1)
\varphi _2 (x,\bar \varepsilon )} \right|$, obtained within the
framework of proposed model, is presented in
Fig.~\ref{susy-article-all-fig:1}. Changes of the   $U_ + ^
+ (\xi ,\lambda )$ shape due to varying $\lambda $ become more and more sharp under $\lambda \to - 1$. The analysis of the expressions of
$U_ + ^ + (\xi ,\lambda )$ shows, that several local minima are
possible only when $0 < \bar \varepsilon < \frac{1}{2}$. Variation
of $\bar \varepsilon $  inside this range changes the height of the
barrier, which signif\/icantly increases with $\bar \varepsilon \to
\frac{1}{2}$. Moreover, the shape of the potential can be
changed (especially the  minima locations) by
varying~$\omega$, when switching to natural variables.
To summarize, the existence of $\left( {\omega ,\bar \varepsilon ,\lambda } \right)$
signif\/icantly change the shape of the $U_ + ^ + (x,\lambda )$.

Substituting  the HO wave functions to (\ref{susy-article-sigma-eqs:37}) and using the Mehler formula \cite{susy-article-sigma-ref:38}, we get $P_ + ^ - (\xi ,z;\xi _0 )$:
      \begin{gather}
         P_ + ^ - (\xi ,z;\xi _0 ) = \left({\omega\over\pi D}\right)^{1/2} {e^{-\xi^2/2} \varphi (\xi,\bar \varepsilon ,1)\over e^{-\xi_0^2/2} \varphi (\xi _0 ,\bar \varepsilon ,1)}\frac{{z^{ - \nu } }}{{\sqrt {1 - z^2 } }}e^{ - \frac{{(\xi z - \xi _0 )^2 }}{{(1 - z^2 )}}},
      \label{susy-article-sigma-eqs:41}
      \end{gather}
where $z = e^{ - \omega t} $. As it was noted in the above $P_ + ^ - (\xi ,z;\xi _0 )$ does not have the equilibrium value and could be normalized (the proof of this claim is given in Appendix), i.e.
      \begin{gather*}
        \int_{ - \infty }^{ + \infty } {d\xi }\; P_ + ^ - (\xi ,z;\xi _0 ) = 1.
      \end{gather*}
The probability density $P_ - ^ - (\xi ,z;\xi _0 )$ is characterized by the existence of the equilibrium value, which is determined by the zero energy wave function of $H_ - ^ - $, i.e.
\[
      P_ - ^ - (x,t \to \infty ;x_0 ) \to \frac{{N^{ - 2} }}{{\varphi ^2 (x,\bar \varepsilon ,1)}},\qquad
      N^{ - 2} = \frac{{2\sqrt \pi }}{{\Gamma ( - \nu )}}.
\]
After some calculations (see Appendix) we obtain the expression for $P_ - ^ - (\xi ,z;\xi _0 )$:
      \begin{gather}
          P_ - ^ - (\xi ,z,\xi _0 ) = \frac{{2\sqrt \pi }}{{\Gamma ( - \nu )\varphi ^2 (\xi ,\bar \varepsilon ,1)}}
 + \frac{1}{{2D^2 }}\left( {\frac{\omega }{{D\pi }}} \right)^{1/2} \! \frac{d}{{d\xi }}\frac{1}{{\varphi (\xi ,\bar \varepsilon ,1)}}\int_0^z\! {d\tau \tau ^{ - (\frac{1}{2} + \bar \varepsilon )} } \Phi (\xi ,\xi _0 ,\tau ),\!\!
      \label{susy-article-sigma-eqs:43a}
\\
          \Phi (\xi ,\xi _0 ,\tau ) = \left( {\frac{d}{{d\xi _0 }}\left[ {F(\xi ,\xi _0 ,\tau )} \right]\varphi (\xi_0 ,\tilde \varepsilon ,1) - F(\xi ,\xi _0 ,\tau )\frac{d}{{d\xi _0 }}\left[ {\varphi (\xi _0 ,\bar \varepsilon ,1)} \right]} \right),
    \nonumber  
\\
        F(\xi ,\xi _0 ,\tau ) = \exp \left( {\frac{{2\xi \xi _0 }}{{1 - \tau ^2 }} - \frac{{(\xi ^2 + \xi _0^2 )}}{2}  \frac{{1 + \tau ^2 }}{{1 - \tau ^2 }}} \right).\nonumber
      \end{gather}
The probability density $P_ - ^ + (\xi ,z;\xi _0 )$ can be obtained
from $P_ + ^ - (\xi ,z;\xi _0 )$ (\ref{susy-article-sigma-eqs:41})
by substituting $\varphi (x,\bar \varepsilon ,c = 1)
 \to \varphi (x,\bar \varepsilon ,\lambda + 1)$. Note that the
normalization condition does not eliminate the $\lambda $-freedom
(see Appendix). As it has been noted in previous section, the expression
for $P_ + ^ + (\xi ,z;\xi _0 )$ is obtained from $P_ - ^ - (\xi
,z;\xi _0 )$  with replacing $  \varphi (x,\bar
\varepsilon ,c = 1) \to \varphi (x,\bar \varepsilon ,\lambda +
1)$, $ N^{ - 2} \to N_\lambda ^{ - 2} =
\frac{{2(\lambda + 1)\sqrt \pi }}{{\Gamma ( - \nu )}}$.

\begin{figure}[t]\centering
          \begin{tabular}{ccc}
          \includegraphics[width=0.35\linewidth]{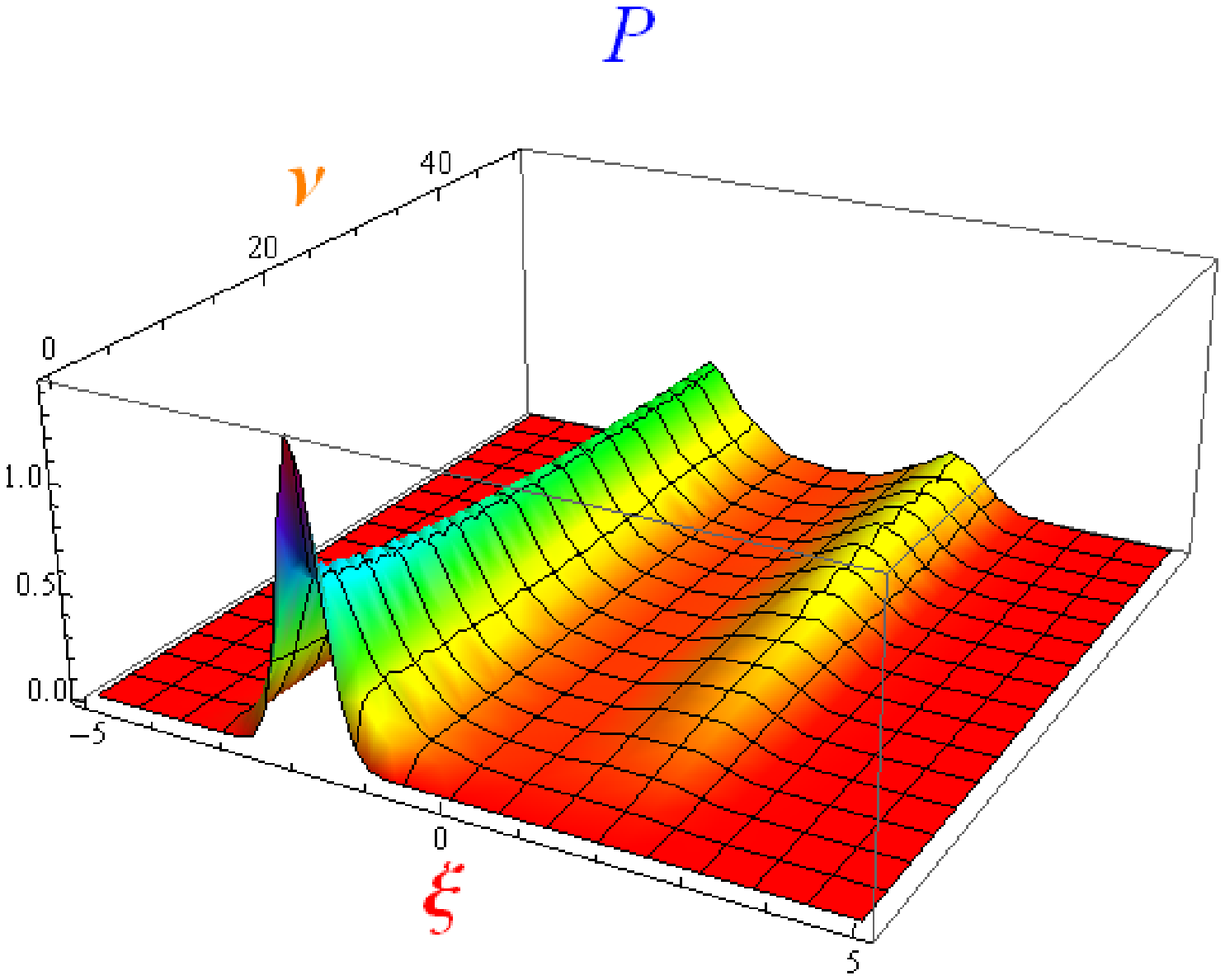} &\qquad \qquad &  \includegraphics[width=0.35\linewidth]{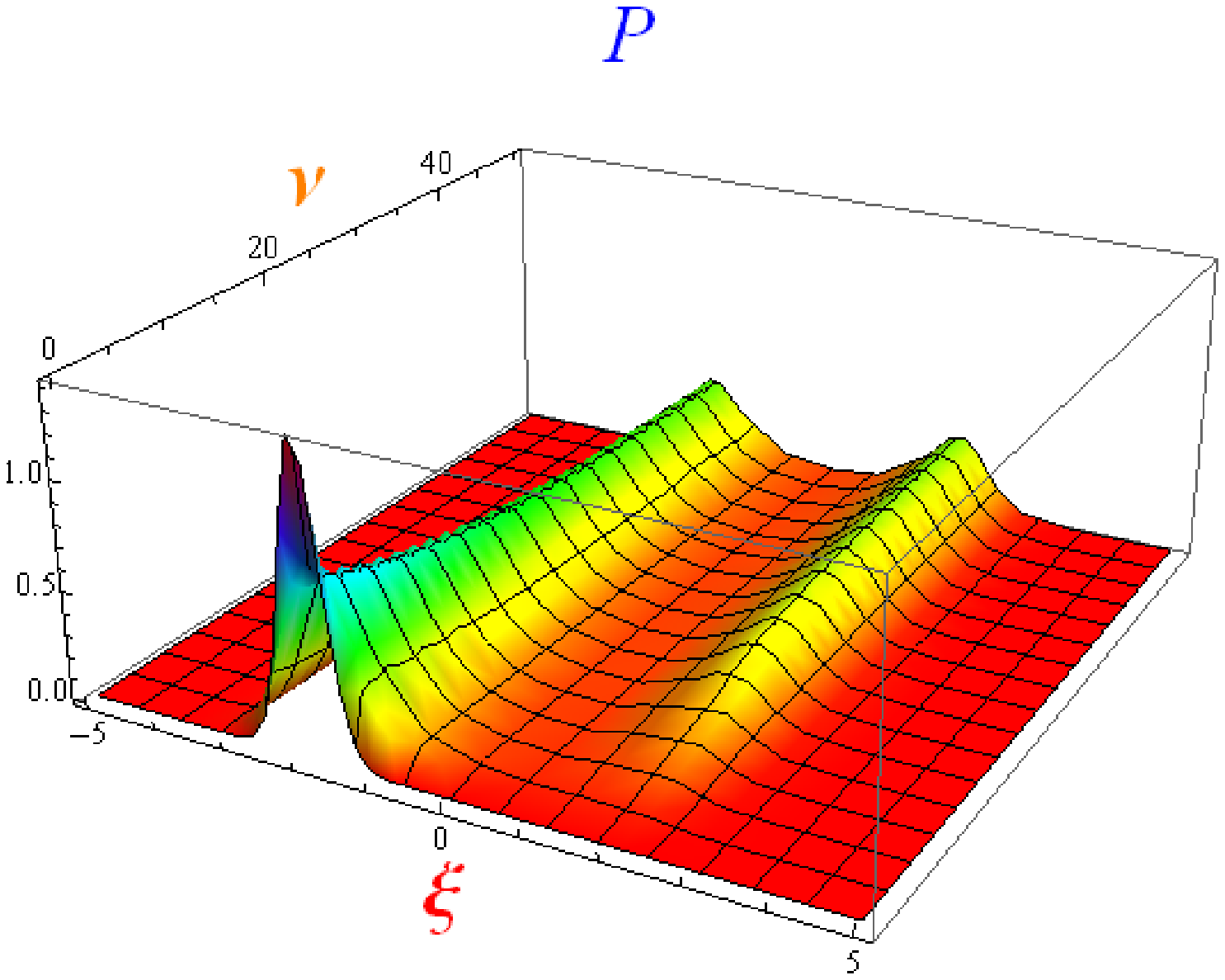}\\
          {\small $\lambda = 0$} &&  {\small $\lambda = 0.5$} \\
          \includegraphics[width=0.35\linewidth]{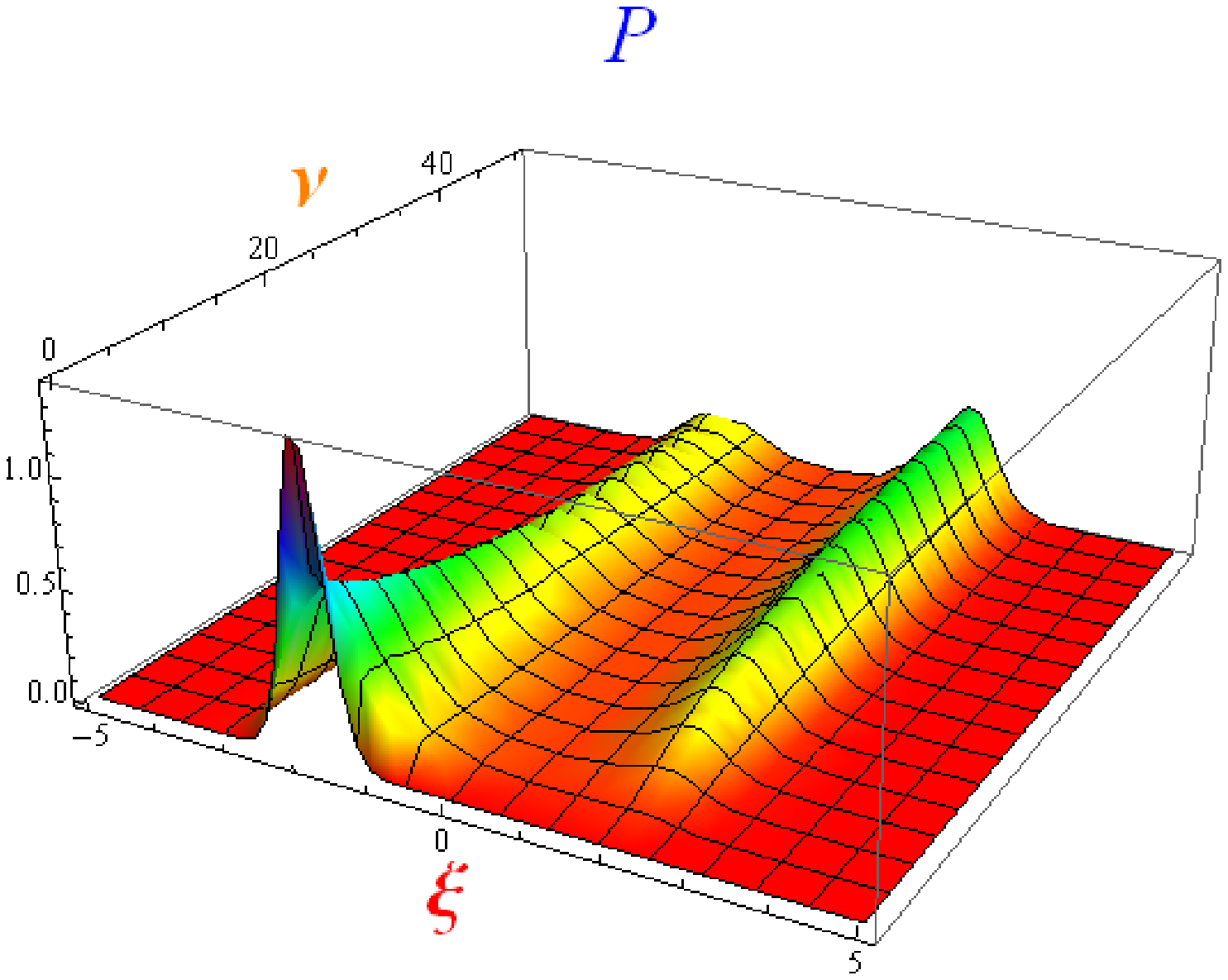} & & \includegraphics[width=0.35\linewidth]{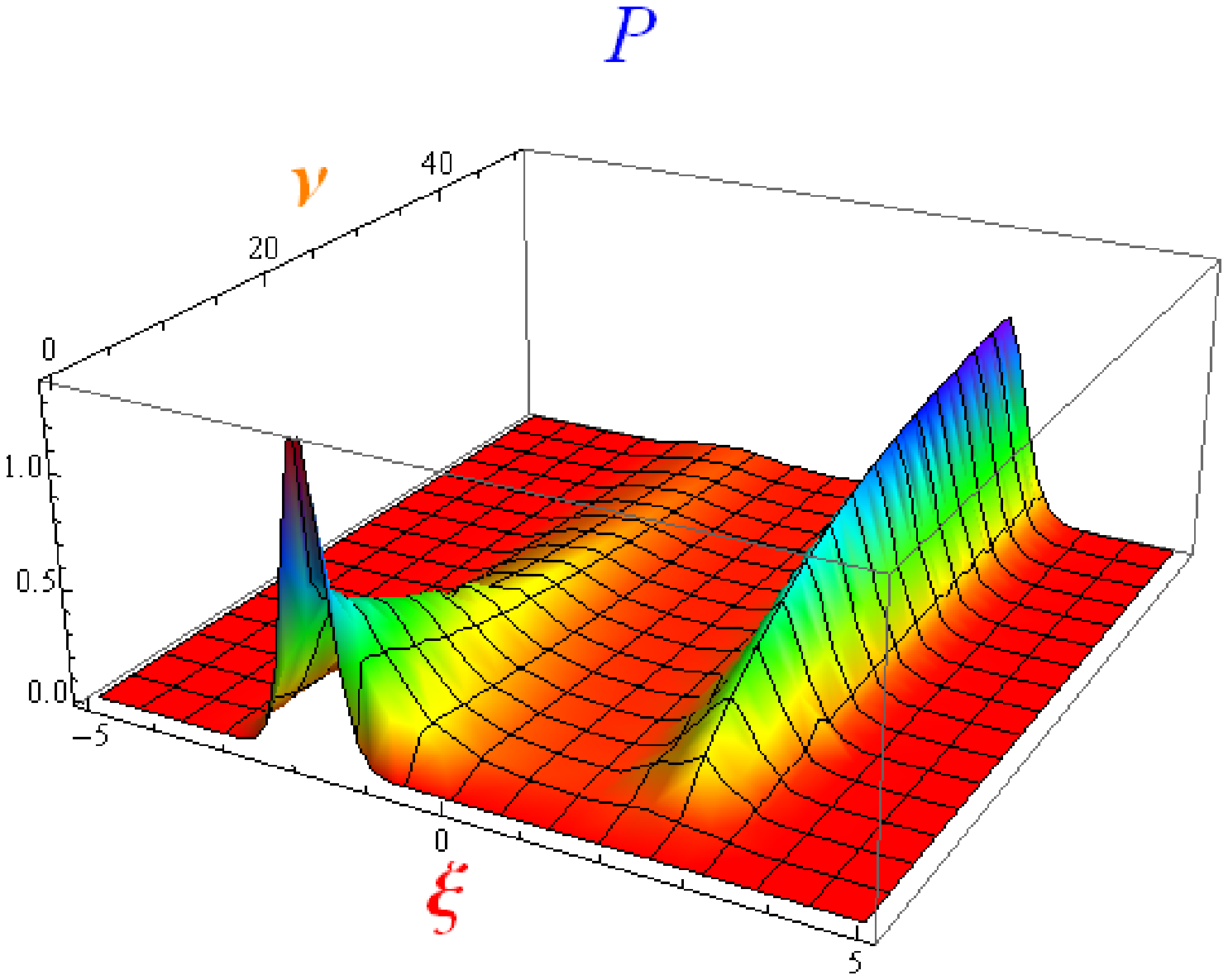}\\
          {\small $\lambda = -0.5$}  &&  {\small $\lambda = -0.95$}
          \end{tabular}
        \caption{$P_ + ^ + (\xi ,z;\xi _0 )$ as a function of $\xi $ and $z$ ($\bar \varepsilon = 0.47$, $\lambda =0$, $\lambda =0.5$, $\lambda =
        -0.5$, $\lambda =-0.95$, $\omega = 1$); $z=e^{-\nu}$.\label{susy-article-all-fig:2}}
      \end{figure}

The dependencies of $P_ + ^ + (\xi ,z;\xi _0 )$ on spatial and time variables, presented in Fig.~\ref{susy-article-all-fig:2}, demonstrate sharp modif\/ications (with f\/ixed $\bar \varepsilon$, $\omega$, $D$) at dif\/ferent values of $\lambda $, especially with $\lambda \to - 1$, when the considerable assymmetry of $P_ + ^ + (\xi ,z;\xi _0 )$ as well as decreasing the time of passing to bimodality are observed. Moreover, modif\/ications of the shape of potential $U_ + ^ + (\xi ,\lambda )$, as well as $P_ + ^ + (\xi ,z;\xi _0 )$, can take place with variation of $\omega $, when we pass from dimensionless variables $\xi $ to physical $x$. This leads to the shift of the local minima and change the height of the barriers. When $\bar \varepsilon \to \frac{1}{2}$, the barrier between local minima signif\/icantly increases, that allows to study the transition time from one local minimum to another as a function of $ \frac{{\Delta U}}{D}$.

\textbf{b)} \textbf{Partial $\boldsymbol{N=4}$~SUSY  breaking.}
Let's brief\/ly discuss partial $N=4$~SUSY breaking and  its consequences to stochastic dynamics. In this case three models for the description of the stochastic dynamics of particles in metastable states ($U_ + ^ - (\xi ,\bar \varepsilon )$, $U_ - ^ - (\xi ,\bar \varepsilon )$, $U_ - ^ + (\xi ,\bar \varepsilon ,\lambda )$) and one   in the bistable state ($U_ + ^ + (\xi ,\bar \varepsilon ,\lambda )$
) arise in the considered approach. The corresponding potentials have the form:
\begin{gather*}
      U_ - ^ - (\xi ,\bar \varepsilon ) = - U_ + ^ - (\xi ,\bar \varepsilon ) = D\ln \big| {D_\nu (\sqrt 2 \xi )} \big|,\\
      U_ + ^ + (\xi ,\bar \varepsilon ,\lambda ) = - U_ - ^ + (\xi ,\bar \varepsilon ,\lambda ) = D\ln \big| {D_\nu (\sqrt 2 \xi ) + \lambda D_\nu ( - \sqrt 2 \xi )} \big|.
\end{gather*}

The range of $\lambda $ variation in $U_ + ^ + (\xi ,\bar \varepsilon ,\lambda )$
and $U_ - ^ + (\xi ,\bar \varepsilon, \lambda)$ is restricted to be $\lambda > 0$. Fig.~\ref{susy-article-all-fig:1}b presents the dependencies of the potentials on spatial variables at f\/ixed $\bar \varepsilon $ and $\lambda $. Stochastic dynamics in mentioned metastable states satisf\/ies to dif\/ferent boundary conditions: ref\/lective boundary at left (right) and absorbing at right (left) for $U_ - ^ - (\xi ,\bar \varepsilon )$ ($U_ + ^ - (\xi ,\bar \varepsilon )$) and absorbing boundaries at both sides for $U_ - ^ + (\xi ,\bar \varepsilon ,\lambda )$. The corresponding probability densities can be calculated with taking into account previously obtained expressions with appropriate modif\/ications.

\section{Conclusions}\label{section7}

In the paper we consider the construction and the study, within the framework of $N = 4$ SUSY~QM, of the
general properties of Hamiltonians with multi-well potentials. These results have been used to obtain
exactly solvable models of stochastic dynamics. The special attention was made to studies of features of these
Hamiltonians without concretization of the form of initial
Hamiltonian. Relations for certain type of integrals,
containing the fundamental solutions to the Schr\"odinger type
equations, allows one to show the form-invariance of the isospectral
Hamiltonians with multi-well potentials, obtained within
of $N = 4$~SUSY~QM. In other words, having the identical spectra of $H_-^-$ and
$H_+^+$, the corresponding potentials can be obtained from each
other through replacing parameters of the problem. Moreover, this
relation allows to analytically calculate the normalization constants
of zero modes, using only asymptotic values of fundamental
solutions. The construction of isospectral Hamiltonians
within $N = 4$~SUSY~QM also implies the partial
supersymmetry breaking. Taking the model of harmonic oscillator as an example we
derive the exact form of the isospectral Hamiltonians with multi-well
potentials and the corresponding wave functions. The existence of a~parameteric freedom gives a possibility to vary the shape of the potential in the wide range.
This becomes important in the research of dif\/ferent phenomena, such as
tunneling processes, which are sensitive to the structure of the multi-well
potentials.

The obtained results have been used for the construction of exactly-solvable
stochastic models. Obtained potentials, entering the Langevin
equations, and probability functions are characteri\-zed by
the parametric dependence, which allows suf\/f\/iciently modify their shape. A~parametric freedom is   typical for
 quantum mechanical isospectral Hamiltonians, where
dif\/ferent versions of the inverse scattering problem  are used
\cite{susy-article-sigma-ref:32, susy-article-sigma-ref:33,
susy-article-sigma-ref:34}. Such a freedom arise in view of an ambiguity
reconstructing the potential from the spectral data. On the other hand, the parametric freedom
in stochastic models  is quite surprising, since their potentials, in contrast to the quantum mechanics, have direct physical meaning. Many characteristics of stochastic dynamics,
such as potential peak passage times and the metastable state lifetime
\cite{susy-article-sigma-ref:35,susy-article-sigma-ref:36,
susy-article-sigma-ref:37}, substantially depend on the shape of
the potential. Therefore, it is interesting to
investigate the characteristics of stochastic processes in
multi-well potentials, such as the Kramers problem, the stochastic
resonance etc., and their dependence on a~modif\/ication of
the shape of the potential. We note that our results admit a generalization to non-Markovian processes, by use of the approach of~\cite{susy-article-sigma-ref:41}. Furthermore, the obtained
results can be generalized to the stochastic models
of polymer dynamics~\cite{susy-article-sigma-ref:42}.
 We have also considered non-trivial consequences of the partial supersymmetry
breaking in $N=4$~SUSY~QM to the description of stochastic
dynamics. In this case three probability density functions with
the same time dependence (i.e.\ with the identical spectrum of
the Fokker--Planck operator) describe stochastic dynamics of particles
in metastable states, and one function corresponds to the dynamics
in the bistable state.

\appendix

\section{Appendix}\label{appendixA}

Let's consider   details of   obtaining the expression of $P_ - ^ - (\xi ,z;\xi _0 )$ and $P_ + ^ + (\xi ,z;\xi _0 )$. According to~(\ref{susy-article-sigma-eqs:36}) the distribution function $P_ - ^ - (\xi ,z;\xi _0 )$ has the form:
      \begin{gather}
          P_ - ^ - (x,t;x_0 ) = \frac{{N^{ - 2} }}{{\left( {\varphi (x,\varepsilon ,1)} \right)^2 }} + \frac{1}{2}\left( {\varphi (x_0 ,\varepsilon ,1)} \right)^2 \nonumber \\
\phantom{P_ - ^ - (x,t;x_0 ) =}{}
\times \sum\limits_{n = 0}^N {\frac{{e^{ - \frac{t}{D}(E_n - \varepsilon )} }}{{(E_n -      \varepsilon )}}} \frac{{d^2 }}{{dxdx_0 }}\left( {\frac{{\psi _ + ^ - (x,E_n )}}{{\varphi (x,\varepsilon ,1)}}} \right)\left( {\frac{{\psi _ + ^ - (x_0 ,E_n )}}{{\varphi (x_0 ,\varepsilon ,1)}}} \right).
        \label{susy-article-sigma-eqs:a1}
      \end{gather}
For the generalized Ornstein--Uhlenbeck process, substituting (\ref{susy-article-sigma-eqs:43})
and the wave function of harmonic oscillator $ \psi _ + ^ - (\xi ,E_n ) = \left( {\frac{\omega }{{\pi D}}} \right)^{1/4} \frac{{D_n (\sqrt 2 \xi )}}{{\sqrt {n!} }}$ into (\ref{susy-article-sigma-eqs:a1}) we get the expression:
      \begin{gather*}
  P_ - ^ - (x,t;x_0 ) = \frac{{N^2 }}{{\varphi ^2 (\xi ,\bar \varepsilon ,1)}} + \frac{1}{{2D^2 }}\left( {\frac{\omega }{{\pi D}}} \right)^{1/2} \frac{d}{{d\xi }}\frac{1}{{\varphi (\xi ,\bar \varepsilon ,1)}}\sum\limits_{n = 0}^\infty {\frac{{z^{n + \frac{1}{2} - \bar \varepsilon } }}{{(n + \frac{1}{2} - \bar \varepsilon ) n!}}} \nonumber \\
 \phantom{P_ - ^ - (x,t;x_0 ) =}{}\times D_n (\sqrt 2 \xi )\;\left( {\frac{d}{{d\xi _0 }}D_n (\sqrt 2 \xi _0 )
 \varphi (\xi _0 ,\bar \varepsilon ,1) - D_n (\sqrt 2 \xi _0 )\frac{d}{{d\xi _0 }}\varphi (\xi _0 ,\bar \varepsilon ,1)} \right),
      \end{gather*}
where $\xi = \sqrt {\frac{\omega }{D}} x$. Using the Mehler formula \cite{susy-article-sigma-ref:38}:
      \begin{gather*}
        \sum\limits_{n = 0}^\infty {\frac{1}{{n!}}} D_n (x)D_n (y)z^n = \frac{1}{{\sqrt {1 - z^2 } }}\exp \left\{ {\frac{{xyz}}{{1 - z^2 }} - \frac{{x^2 + y^2 }}{4} \frac{{1 + z^2 }}{{1 - z^2 }}} \right\} \equiv F(x,y;z)
      \end{gather*}
and the relation
      \begin{gather*}
        \sum\limits_{n = 0}^\infty {\frac{{z^{n + \frac{1}{2} - \bar \varepsilon } }}{{n!\left( {n + \frac{1}{2} - \bar \varepsilon } \right)}}} D_n (x)D_n (y)  = \int_0^z {d\tau \;\tau ^{n - \left( {\frac{1}{2} + \bar \varepsilon } \right)} } F(x,y,\tau )
      \end{gather*}
we obtain the expression for $P_ - ^ - (\xi ,z;\xi _0 )$ (\ref{susy-article-sigma-eqs:43a}). It should be noted that presence of spatial derivatives eliminate the contribution from excited states of the Hamiltonian $H_ - ^ - $ to the normalization condition for $P_ - ^ - (\xi ,z;\xi _0 )$.

Let's consider the question of normalizability of $P_ + ^ - (x,t;x_0 )$ and $P_ - ^ + (x,t;x_0 )$ as well as the possibility of elimination of the $\lambda $-freedom in $P_ - ^ + (x,t;x_0 )$ by the normalization condition. As it was noted  in the above, $P_ - ^ + (x,t;x_0 )$ is obtained from $P_ + ^ - (x,t;x_0 )$ by replacing $\varphi (x,\bar \varepsilon ,1) \to \varphi (x,\bar \varepsilon ,\lambda + 1)$ and has a form
      \begin{gather*}
        P_ - ^ + (\xi ,z;\xi _0 ) = \left( {\frac{\omega }{{\pi D}}} \right)^{1/2} \frac{{e^{-\xi^2/2} \varphi (\xi ,\bar \varepsilon ,\lambda + 1)}}{{e^{-\xi_0^2/2} \varphi (\xi _0 ,\bar \varepsilon ,\lambda + 1)}}\frac{{z^{ - \nu } }}{{\sqrt {1 - z^2 } }}e^{ - \frac{{(\xi z - \xi _0 )^2 }}{{(1 - z^2 )}}}.
      \end{gather*}
The normalization condition for the distribution function:
      \begin{gather*}
        \int_{ - \infty }^{ + \infty } {dx\;P_ - ^ + (x,t;x_0 )} = \sqrt {\frac{D}{\omega }} \int_{ - \infty }^{ + \infty } {d\xi \; P_ - ^ + (\xi ,z;\xi _0 )} \equiv 1.
      \end{gather*}
Substituting the integral representation of the parabolic cylinder function
      \begin{gather*}
        D_p (z) = \frac{{e^{ - \frac{{z^2 }}{4}} }}{{\Gamma ( - p)}}\int_0^\infty {e^{ - zx - \frac{{x^2 }}{2}} }   x^{ - p - 1}dx ,\qquad {\rm Re}\, p < 0,
      \end{gather*}
results in
      \begin{gather}
        \int_{ - \infty }^{ + \infty } {d\xi \;} e^{ - \frac{{(\xi z - \xi _0 )^2 }}{{(1 - z^2 )}}}  e^{-\xi^2/2} \;D_\nu ( \pm \sqrt 2 \xi ) = \sqrt \pi \;z^\nu (1 - z^2 )^{1/2}\;e^{-\xi_0^2/2}  D_\nu ( \pm \sqrt 2 \xi _0 ).
      \label{susy-article-sigma-eqs:a8}
      \end{gather}
Substituting (\ref{susy-article-sigma-eqs:a8}) to the normalization condition for $P_ - ^ + (x,t;x_0 )$ and $P_ + ^ - (x,t;x_0 )$ it is easy to verify, that they are equal to one, thus the probability densities, which correspond to the stochastic dynamics in a~metastable state, are normalizable and the normalization condition does not eliminate the $\lambda $-freedom in $P_ - ^ + (x,t;x_0 )$.

\subsection*{Acknowledgements}

Authors thank to M.~Plyushchay for helpful discussions and BVP Conference Organizers for a~stimulating environment. We are thankful to A.~Nurmagambetov for reading the manuscript, suggestions and improvings.

\pdfbookmark[1]{References}{ref}
\LastPageEnding

\end{document}